\documentclass[journal,twoside,web]{ieeecolor}
\usepackage{tmi}
\usepackage{cite}
\usepackage{amsmath,amssymb,amsfonts}
\usepackage{algorithmic}
\usepackage{graphicx}
\usepackage{textcomp}
\usepackage[dvipsnames]{xcolor}
\usepackage{stackengine}
\usepackage{hyperref}
\usepackage{textcomp}
\usepackage{slashbox}
\usepackage{caption}
\newcommand{\reals}{I \! \! R}
\usepackage{algorithm,algorithmic}
\usepackage{lipsum}
\usepackage{mathtools}
\usepackage{cuted}
\usepackage{footnote}
\usepackage{textcomp}
\usepackage{algorithmic}
\usepackage{verbatim}
\usepackage{multirow}
\usepackage{subfig}
\usepackage{tablefootnote}
\usepackage{url}
\usepackage[flushleft]{threeparttable}
\newcommand{\norm}[1]{\left\lVert#1\right\rVert}

\newcommand{\RobRevision}[1]{\textcolor{black}{#1}}

\def\BibTeX{{\rm B\kern-.05em{\sc i\kern-.025em b}\kern-.08em
   T\kern-.1667em\lower.7ex\hbox{E}\kern-.125emX}}
\newcommand{\TMIRevision}[1]{\textcolor{black}{#1}}   
\newcommand{\TMISecondRevision}[1]{\textcolor{black}{#1}}
\begin{document}
\title{Multifrequency 3D Elasticity Reconstruction with Structured Sparsity and ADMM}
\author{Shahed~Mohammed, Mohammad~Honarvar, Qi~Zeng, Hoda~Hashemi,Robert~Rohling, Piotr~Kozlowski, Septimiu~Salcudean 
\thanks{This work was funded by NSERC, CIHR and the Charles Laszlo Chair in Biomedical
Engineering held by Professor Salcudean.}
\thanks{S. Mohammed, Q.~Zeng, H.~Hashemi, R.~Rohling, S.~Salcudean are with the Department of Electrical and Computer Engineering, University of British Columbia, BC. (email: shahedkm@ece.ubc.ca, rohling@ece.ubc.ca, and tims@ece.ubc.ca). Dr. M. Honarvar was with the same department. (email: honarvar@ece.ubc.ca)}
\thanks{\RobRevision{R. Rohling is also with the Department of Mechanical Engineering, University of British Columbia.}}
\thanks{P. Kozlowski is with the Department of Radiology, University of British Columbia, BC. (email: piotr.kozlowski@ubc.ca)}}
\maketitle

\begin{abstract}
We introduce a model-based iterative method to obtain shear modulus images of tissue using \TMIRevision{magnetic resonance elastography}. The method jointly finds the displacement field that best fits multifrequency tissue displacement data and the corresponding shear modulus.  The displacement satisfies a viscoelastic wave equation constraint, discretized using the finite element method. Sparsifying regularization terms in both shear modulus and the displacement are used in the cost function minimized for the best fit.
The formulated problem is bi-convex. Its solution can be obtained iteratively by using the alternating direction method of multipliers.
\TMIRevision{Sparsifying regularizations and the wave equation constraint filter out sensor noise and compressional waves.} Our method does not require bandpass filtering as a preprocessing step and converges fast irrespective of the initialization. We evaluate our new method in multiple \TMIRevision{{\em in silico}} and phantom experiments, with comparisons with existing methods, and we show improvements in contrast to noise and signal to noise ratios. \TMISecondRevision{Results from an \emph{in vivo} liver imaging study show elastograms with mean elasticity comparable to other values reported in the literature.}   
\end{abstract}


\section{Introduction}
\IEEEPARstart{E}{lastography} is a non-invasive approach to image the tissue's mechanical properties, such as elasticity and viscosity \cite{Ophir1991}, which can provide valuable information on structural change caused by different diseases ~\cite{Hirsch2013, Murphy2018, Sebastian2017_Book}. Elastography reconstruction~(ER) involves solving a non-linear ill-posed inverse problem to retrieve the tissue elasticity from displacement captured by magnetic resonance imaging (MRI) or ultrasound (US)~\cite{Honarvar2012}. In order to increase the numerical stability of elasticity reconstruction, the tissue is often considered to be locally homogeneous and incompressible, with tissue waves assumed to be purely in shear~\cite{Papazoglou2012}. These assumptions give rise to an independent Helmholtz equation for each component of the displacement, leading to fast and straightforward reconstruction methods such as Direct Inversion~\cite{Oliphant2001a}, local frequency estimation (LFE)~\cite{Manduca1996}, and  tomoelastography (k-MDEV)~\cite{Tzschatzsch2016}. The local homogeneity assumption is not met in most tissues. Therefore, when a homogeneous model is used in the ER, significant boundary artifacts and amplification of measurement noise may appear, reducing the reconstructed elasticity map's resolution and quality~\cite{Honarvar2016}. \par

\TMISecondRevision{Moreover, ER is an ill-posed inverse problem. As a result, removing noise effects requires careful filter design as a preprocessing stage~\cite{Barnhill2017}.}  Most approaches utilize tissue displacement bandpass filtering as a preprocessing stage to constrain the resultant elasticity to a known range~\cite{Papazoglou2012}. However, designing a bandpass filter for a narrow range of elasticity without losing relevant information can be difficult.  
As a result, bandpass filtering either smoothes out the tissue boundaries or amplifies the boundary artifacts. 
\TMIRevision{Moreover, the effectiveness of bandpass filtering in removing the compressional waves is still under question for heterogeneous media. 
Therefore, numerical derivative methods such as curl-based filtering of the displacements are a requirement for extracting the shear waveform~\cite{baghani2009theoretical}. 
However, numerical derivative methods  tend to amplify the measurement noise in the displacement pattern.}   Alternatively, the weighted average of elasticity maps from multifrequency and multi-component displacement are used in techniques such as multifrequency dual elasticity viscosity reconstruction~\cite{Papazoglou2012} and tomoelastography~\cite{Tzschatzsch2016}. These methods use empirically weighted averaging to reduce the frequency-dependent sparse noise and the measurement noise. These methods have shown improved fidelity for simulated and real-tissue data compared to reconstruction from a single displacement field~\cite{Tzschatzsch2016}.\par

\TMISecondRevision{An alternative way to circumvent the denoising stage is to use a  iterative model fitting approach to fit the displacement to the underlying elastic wave equation~\cite{VanHouten1999}.} In contrast to direct inversion, the displacement is calculated using the wave equation, and iteratively the shear modulus that minimizes the distance between the measured displacement and simulated displacement is generated. \TMIRevision{As a result, iterative methods are more robust to data quality and do not require preprocessing, such as bandpass filtering.} However, as the iterative model fitting for ER is a non-convex optimization algorithm, global convergence is not guaranteed, and the reconstruction results strongly depend on the initialization \cite{Otesteanu2018d}. Most of the iterative reconstruction methods in elastography solve the required non-linear problem by utilizing a Newton-like approach \cite{Tan2017}. \RobRevision{In the absence of surface force measurements, the gradient calculations of the iterative methods are sensitive to the conditioning of the displacement fitting, requiring  high mesh density to converge~\cite{Honarvar2016}}. The use of multi-initialization, along with multifrequency modelling can provide more robust reconstruction at the expense of higher computational cost~\cite{Otesteanu2018d}. \par

\TMISecondRevision{Our recent preliminary work has reformulated the 2D iterative ER method as a convex optimization problem with a bi-affine equality constraint, which was solved using the alternative directional method of multipliers (ADMM) ~\cite{Mohammed2019}}. Termed as \textbf{E}lasticity \textbf{R}econstruction using \textbf{B}i-convex \textbf{A}DMM (ERBA), \RobRevision{this method allowed the recovery of both the elasticity and displacement fields using a series of direct solutions of the displacement fitting and elasticity inversion, where both problems are well-conditioned.} The results have shown better reconstruction quality with robustness to initialization. Additionally, the use of the wave \TMIRevision{equation} as an equality constraint proved to provide better results for stiff inclusions than using the wave constraint as a penalty as in \cite{Otesteanu2018}. However, as a two-dimensional method, ERBA requires the plane stress assumption for an accurate estimation of shear modulus. 
\TMIRevision{Studies of the waveguide effect have shown that 3D inversion and curl-filtering are required for accurate estimation of elasticity in heterogeneous materials~\cite{baghani2009theoretical }.   Moreover, frequency-dependent sparse noise associated with displacement nodes generates artifacts in the image reconstruction, which are hard to eliminate with single frequency displacement fields~\cite{Papazoglou2012}. }\par
\vspace{-0.2in}
\subsection{Paper Contribution}
\label{subsec:: Paper Contribution}
\TMISecondRevision{This paper reports two elasticity reconstruction methods termed \textbf{E}lasticity \textbf{R}econstruction with dual \textbf{S}parsity by \textbf{A}DMM (ERSA) and \textbf{M}ultifreqeuncy \textbf{E}lasticity \textbf{R}econstruction with dual \textbf{S}parsity by \textbf{A}DMM (MERSA).  
Unlike ERBA, ERSA and MERSA uses a 3D viscoelastic wave constraint, and is formulated using a mixed finite element model (FEM) that takes the hydrostatic pressure as an additional unknown variable~\cite{park2006shear, Honarvar2012}. In addition, MERSA uses a 3D multifrequency viscoelastic wave constraint to fit multi-frequency displacements with a single elasticity map. This model does not require the local homogeneity assumption and does not neglect the compressional waves. As a result, the mixed-FEM can provide greater accuracy and resolution~\cite{Honarvar2017comparison}. This paper also presents a mixed-FEM system for multifrequency displacement data.} While multifrequency iterative ER~\cite{Otesteanu2018d} and 3D iterative ER~\cite{Tan2017} have been proposed in previous literature,  MERSA incorporates multifrequency and works with 3D displacements. As a result, MERSA does not require plane stress or plane strain assumptions and can additionally provide high quality ER while eliminating frequency-dependent sparse noise. \par

\TMISecondRevision{Previous multifrequency iterative ER methods have shown a strong dependency on the initialization due to the convergence being only local~\cite{Otesteanu2018d}}. To improve convergence, we incorporate in MERSA sparsity priors on both displacement and elasticity. \TMIRevision{Particularly, we propose a k-space sparsity prior that filters out the high frequency dense sensor noise.} Moreover, we apply an isotropic total variation (TV) prior on the elasticity to retain sharp edges. Although sparsity of measurements has been previously shown to provide better performance for low dose computed tomography reconstruction \cite{He2019}, such approaches have not been investigated in ER previously.  \par 

We also incorporate a sub-zone based method similar to \cite{VanHouten1999, Tan2017} to reduce computational complexity. Unlike previous methods, we apply the global regularization prior on the elasticity map, which significantly reduces the block artifacts and requires less overlap of the sub-zones. Finally, the use of bi-convex ADMM in MERSA with consensus optimization over the elasticity map allows closed-loop well-conditioned inversion of displacement and elasticity for each sub-zone separately. Using sub-zones reduces the computational burden in the 3D FEM matrix formulation and inversion. The variable splitting method associated with ADMM allows easy implementation of the sparsity regularization on displacement. Besides, the consensus ADMM formulation ensures retaining the separability over the sub-zones while imposing a 3D total variation prior on the global elasticity map.\par
We evaluate the performance of MERSA in comparison with well-established ER methods in numerical simulations, phantom studies, and in an \emph{in vivo}  liver study to demonstrate the goal of achieving high resolution and accuracy with robustness to measurement data quality. \par
\vspace{-0.1in}
\subsection{Paper Organization}
The rest of the paper is organized as follows: The new MERSA method is presented in Section \ref{sec:: Methods}. We detail the experimental setups, datasets, evaluation metrics, and the other methods used for comparison in Section \ref{sec :: Method Evaluation}. The results are described in Section \ref{sec :: Results}, and a detailed discussion of the results is presented in Section~\ref{sec:: Discussion}. 

\section{Methods}
\TMISecondRevision{In this section, we first introduce the conventional FEM elasticity reconstruction using  the direct method and the iterative method. Then, we present our proposed methods for single frequency (ERSA) and multi frequency (MERSA), and the optimization model that is used for reconstruction. }  
\label{sec:: Methods}
%\begin{figure*}[!t]
%	\centering
%	{\includegraphics[clip,trim=0 2cm 0 0,width=0.8\linewidth]{./Figures/Block_Diagram/MERSA.pdf}}	
%	\caption{Overview of MERSA. In each iteration, a new elasticity is inverted using the model displacement ($\mathbf{u}$), which is denoised using TV regularization. This elasticity map is used for recovering the displacement using the forward problem, which is denoised using sparse prior regularization in k-space. In the next iteration, the new displacement is used for retrieving the elasticity map}
%	\label{Cross Validation20}
%\end{figure*}
\subsection{FEM Formulation for Elastic Waves in Tissue}
\label{subsec:Theory}
\TMISecondRevision{Consider a linear, isotropic, incompressible, elastic medium subjected to a time harmonic excitation with a driving frequency $\omega$.  If $\mathbf{u}: \reals^3 \times \reals \rightarrow \mathbb{C}^3$, $ (x , \omega) \rightarrow \mathbf{u}(x,\omega)$ is the discrete displacement field in the Fourier domain, we can discretize the elastic wave equation using the Finite Element Method (FEM) as follows ~\cite{Honarvar2016}:
\begin{equation}
\left[ \mathbf{K}_{\mu} \left( \mu\right)-\omega^2 \mathbf{M}\left(\rho\right)\right]\, \mathbf{u}+ \mathbf{K}_p\, p= \mathbf{0} \,\,\, , \label{Linearized Equation_Mixed}
\end{equation}
where the first term $\mathbf{K}_{\mu}$ is the stiffness matrix, which is a function of the shear modulus $\mu$. The second term  $\mathbf{M}$ is the mass matrix, which is a function of the density of the material ($\rho$). The third term is associated with the pressure variable defined by $p=\lambda \left(\nabla \cdot \mathbf{u}\right)$. Here, $\lambda$ is the Lam\'{e}'s first parameter, and $\nabla$ is the gradient operator. The model in Eq. \ref{Linearized Equation_Mixed} is called the ``mixed-FEM" \cite{park2006shear, Honarvar2012}. In this model, the complex shear modulus captures both the storage shear modulus, and the loss shear modulus as the real and imaginary parts of $\mu = \mu_r + i \mu_i$. The complex shear modulus $\mu$ can be reconstructed using either direct inversion or iterative methods from the mixed-FEM model.} 
\subsubsection{Mixed-FEM Direct Inversion of Shear Modulus}
%{We assume that $\mu$ does not vary with frequency because we use an acquisition protocol with little separation between excitation frequencies.}
%{where $\rho$ is the (assumed constant) density of the material, $\lambda$ is the Lam\'{e} first parameter, and $\nabla$ is the gradient operator. }

% TS I DO NOT THINK YOU NEED THIS If $\nu$ is the Poisson's ratio of the medium, then the shear modulus $\mu$ and Lam\'{e} first parameter $\lambda$ are related to the Young's Modulus $E$ through the following equations:
%\begin{align}
%    \mu= \frac{E}{2\left(1+\nu\right)} \,\,\,\,\,\,&,& \lambda= \frac{\nu E}{\left(1+\nu\right)  \left(1-2\nu\right)}
%\end{align}
\par 

\TMIRevision{Reconstructing the shear modulus using the direct inversion method involves reorganizing the terms in equation (\ref{Linearized Equation_Mixed}) to obtain linearized equations for $\mu$ and $p$ given by \cite{Honarvar2012, park2006shear}:}
\TMIRevision{
\begin{equation}
\mathbf{K}_u \left(\mathbf{u}\right) \, \mathbf{\mu}+ \mathbf{K}_p\,p =\mathbf{f}\left(\mathbf{u}\right) \label{Linearized Equation_IP} \,\,\, 
\end{equation}
where $\mathbf{K}_u$ is the resultant coefficient matrix after reorganizing the first term in equation (\ref{Linearized Equation_Mixed}) with respect to $\mu$, with $\mathbf{K}_u\left(\mathbf{u}\right)\mathbf{\mu}= \mathbf{K}_\mu \left( \mu\right)\mathbf{u} $ and   $\mathbf{f}\left(\mathbf{u}\right) =\omega^2 \mathbf{M}\mathbf{u} $}. \TMIRevision{ Direct inversion methods use the measured displacement directly in the wave constraint model to reconstruct the elasticity. Therefore, any inconsistency with the wave constraint model and the real displacement can cause artifacts in the reconstructed elastogram. Direct inversion methods usually involve a displacement preprocessing stage such as bandpass filtering to remove measurement noise and compressional waves from the measured displacement pattern. Moreover, regularization techniques such as sparsity regularization~\cite{Honarvar2012} are necessary for stable inversion.  }  
%\vspace{-0.15in}
\subsubsection{Mixed-FEM Iterative Methods}
In iterative methods for elasticity reconstruction~\cite{VanHouten1999,Tan2017, Otesteanu2018}, the ER problem is formulated as a constrained optimization problem that finds the shear modulus distribution ${\mu}^*$ \TMIRevision{and pressure variable ${p}^*$} that minimize the distance between the measured displacement $\mathbf{v}$ and the calculated displacement phasor $\mathbf{u}$ from (\ref{Linearized Equation_Mixed}):
\begin{eqnarray}
\label{Iterative_OP}
\min_{\begin{array}{cl}\mathbf{\mu} \in \mathbb{C}^{N} \\ {p} \in \mathbb{C}^{N} \end{array}} \hspace{-15pt}
&& \Bigg\{ \frac{1}{2} \norm{ \mathbf{u}\left(\mu,p\right)-\mathbf{v}}^2+ \mathcal{R}\left(\mu\right) \Bigg|
\nonumber \\[-20pt] 
&& \begin{array}{cl}
\left[ \mathbf{K}_\mu\left({\mu}\right)-\omega^2 \mathbf{M}\right] \, \mathbf{u}+ \mathbf{K}_pp= \mathbf{0} & \text{in } \Omega  \\
p=\lambda \left(\nabla \cdot \mathbf{u}\right)
\end{array}
\Bigg\},
\end{eqnarray}
where $\mathcal{R}\left(\mu\right)$ represents a regularization function to apply {\em a priori}  knowledge on $\mu$. \TMIRevision{Iterative methods are more robust to low displacement quality as the noisy measured displacement is fitted to the given wave equation constraint. Thus we expect an iterative method to more effectively remove noise in the displacement pattern while reconstructing the elastogram.} However, the wave constraint model constitutes a bi-linear constraint optimization in terms of $\mu$ and $\mathbf{u}$, and therefore does not guarantee global convergence. Most methods solve the optimization using a quasi-Newton or a conjugate gradient method. This requires calculating the Jacobian of the cost function given in (\ref{Iterative_OP}) with respect to the shear modulus $\mu$, which does not have a closed-form solution. Numerical differentiation to find the Jacobian would require solving the forward problem $N$ times for each iteration. Alternatively, the Jacobian can be determined using only two solutions of the forward problem for each iteration using the adjoint method~\cite{Tan2017}. \TMISecondRevision{In our preliminary work presented as ERBA~\cite{Mohammed2019}, we have introduced bi-convex ADMM as an alternative optimization technique to gradient descent. In the case of the 2D scalar wave equation, ERBA resulted in less dependency on the initialization and more robustness to noise. However, ERBA did not consider the hydrostatic pressure. In the following, we extend ERBA to the 3D mixed-FEM formulation.}   
\vspace{-0.05in}
\begin{comment}

\subsection{ Simultaneous Reconstruction of Displacement and Elasticity}
For high quality elasticity reconstruction, the modelled displacement $\mathbf{u}$ must be accurately estimated from the measured displacement $\mathbf{v}$ without the proper boundary conditions. 
Since $\mu$ and $\mathbf{u}$ are related by the bi-linear constraint function given by (\ref{Linearized Equation_Mixed}), simultaneous reconstruction of $\mu$ and $\mathbf{u}$ can provide robust reconstruction results, as shown in \cite{Otesteanu2018}. Termed as a Hybrid FEM, this method applies (\ref{Linearized Equation_Mixed}) as a penalty function to find $\left({\mu}^*, {\mathbf{u}}^*, p^*\right)$ that solve:
\begin{align}
\label{Hybrid_FEM} 
\underset{\begin{subarray}{c}
	\mathbf{\mu} \in \mathbb{C}^{N}\\
	\mathbf{u} \in \mathbb{C}^{3N}\\
	p \in \mathbb{C}^{N}
	\end{subarray}}{\min} 
	&\Bigg\{ \frac{\rho}{2} \norm{ \mathbf{u}-\mathbf{v}}^2+ \gamma \norm{\mu}_{TV}+ \nonumber  \\[-20pt]  & \,\,\, \,\,\, \frac{1}{2}\norm{\left[ \mathbf{K}_\mu \left( \mu\right)-\omega^2 
	\mathbf{M}\right] \mathbf{u}+ \mathbf{K}_p \, p}^2\Bigg\} 
\end{align}
where the first two terms ensure  data consistency of $\mathbf{u}$, while the last term applies the Total Variation (TV) prior on $\mu$. 
The relative importance of the data fitting and the TV prior are controlled by the weights $\rho$ and $\gamma$.
\vspace{-0.1in}
\end{comment}
\vspace{-0.2in}
\subsection{ERSA: Elasticity Reconstruction using Structured Sparsity and ADMM}
\label{subsec:DualSparsity}
We adopt the following dual-domain sparse reconstruction model as our framework for elastography reconstruction:
\begin{eqnarray}
\label{Iterative_ERBA}
 \underset{\begin{subarray}{c}
	\mathbf{\mu} \in \mathbb{C}^{N}\\
	\mathbf{u} \in \mathbb{C}^{3N}\\
	p \in \mathbb{C}^{N}
	\end{subarray}}{\min} 
 \,\,\,
\Bigg\{ \frac{\rho}{2} \norm{ \mathbf{u}-\mathbf{v}}^2+ \mathcal{R}_{\mu}\left(\mu\right)+ \mathcal{R}_{u}\left(\mathbf{u}\right)+ \mathcal{R}_{p}\left(\mathbf{p}\right) \Bigg| \nonumber \\[-20pt] 
%\mbox{subject to }
\begin{array}{cl}
\left[ \mathbf{K}_\mu \left( \mu\right)-\omega^2 \mathbf{M}\right] \, \mathbf{u}+\mathbf{K}_p \,p= \mathbf{0} & \text{in } \Omega \\
\end{array} 
\Bigg\}
\end{eqnarray}
\TMIRevision{We termed this formulation as \textbf{E}lasticity \textbf{R}econstruction using \textbf{S}tructured Sparsity and \textbf{A}DMM (ERSA).} In the above equation, the first quadratic error term penalizes the difference between the measured displacement $\mathbf{v}$ and calculated displacement $\mathbf{u}$. The second, third, and fourth terms are the regularization functions for elasticity $\mathcal{R}_{\mu}\left(\mu\right)$, displacement $\mathcal{R}_{u}\left(\mathbf{u}\right)$, and \TMIRevision{hydrostatic pressure $\mathcal{R}_{p}\left(p\right)$} respectively. 
For elasticity, we use an isotropic TV regularizer, which assumes tissue elasticity to be piece-wise smooth. The TV regularizer also contains a box-constraint prior to restrict the elasticity values to a specific range.\par
\TMISecondRevision{The displacement phasor noise can be modelled accurately by a combination of high spatial frequency dense noise and low spatial frequency dense noise~\cite{Barnhill2017}}. Therefore we use a k-space sparsity prior to remove the dense noise from the spatial frequency domain. \TMIRevision{The k-space sparsity prior, the wave constraint fitting, and the box-constraint prior on the elasticity values mimic the effect of bandpass filtering to remove high frequency noise and low frequency compressional waves. The use of box constraints in elasticity estimation further filters out the long wavelength components of compression waves from the measured displacement.} Here, we have used an orthogonal 3D-FFT transform to take the displacement to k-space, where we used a sparsifying $L^1$ norm:
\begin{subequations}
\begin{align}
\mathcal{R}_{\mu}\left(\mu\right)=& \gamma_{\mu} \norm{\mu}_{TV}\\
\mathcal{R}_{u}\left(\mathbf{u}\right)=& \gamma_{u} \norm{\hat{\mathbf{u}}}_1=\gamma_{u} \norm{\text{FFT}\left(\mathbf{u}\right)}_1 \,\,,\\
\mathcal{R}_{p}\left(p\right)=& \frac{\gamma_{p}}{2} \norm{\nabla p}_{2}^2 \,\, ,
\end{align}   
\end{subequations}
\TMISecondRevision{where $\norm{\cdot}_{TV}$} represents the isotropic discretized TV-norm~\cite{Beck2009},   
$\hat{\mathbf{u}}$ is the corresponding k-space transform of $\mathbf{u}$, and $\gamma_\mu$ and $\gamma_u$ are regularization weights. 
\TMIRevision{We used the $L^2$ norm of the hydrostatic pressure gradient to regularize the hydrostatic pressure $p$. The difference between ERSA and ~\cite{Otesteanu2018d} is the use of an exact constraint on the wave equation for simultaneous reconstruction of $\mathbf{u}$ and $\mu$ with dual sparsity, and the bi-convex structure to employ ADMM.} \TMISecondRevision{Next, we describe the extension of ERSA for multifrequency displacement.}\par
\vspace{-0.2in}
\subsection{MERSA:  Multifrequency ERSA}
In harmonic elastography, regions of zero displacement amplitude cause most inversion methods to fail and are difficult to predict as they depend on poorly defined boundary conditions. As a result, most direct inversion methods utilize multifrequency excitation to generate robust elasticity reconstruction~\cite{Papazoglou2012, Tzschatzsch2016}. \TMIRevision{In our model, we use a multifrequency joint reconstruction to find $\left({\mu}^*, {\mathbf{U}}^*, {P}^*\right)$ that solve the analog of Eq. \ref{Linearized Equation_Mixed} for $J$ frequencies:
\begin{align}
\label{Optimization Model_MF}
\underset{\begin{subarray}{c}
	\mathbf{\mu} \in \mathbb{C}^{N}\\
	\mathbf{U} \in \mathbb{C}^{3NJ}\\
	{P} \in \mathbb{C}^{NJ}
	\end{subarray}}{\min} 
 \,\,\,
\Bigg\{
\frac{\rho}{2} \norm{\mathbf{U}-\mathbf{V}}^2+ \mathcal{R}_{\mu}\left(\mu\right)+ \mathcal{R}_{U}\left(\mathbf{U}\right)+ \mathcal{R}_{P}\left(P\right) \nonumber \\[-18pt] 
\Bigg|
\begin{array}{cl}
\left[ \mathbb{K}_{\mu} \left( \mu\right)-\mathbb{M}\right] \, \mathbf{U}+ \mathbb{K}_{P}\,P= \mathbf{0} & \text{in  } \Omega
\end{array}
\Bigg\}
\end{align}
}
\TMISecondRevision{where,  $\mathbf{U}$, $\mathbf{V}$, $P$, $\mathbb{K}$, $\mathbb{M}$, $\mathbb{K}_p$, $\mathcal{R}_{U}$ and $\mathcal{R}_{p}$ are now reperesenting multi-freqeuncy displacements. Their definitions are given in the appendix.}
\vspace{-0.1in}
\subsection{Optimization Using Sub-zone-based Inversion}
\begin{figure}[!t]
	\centering
	{\includegraphics[width=0.45\linewidth]{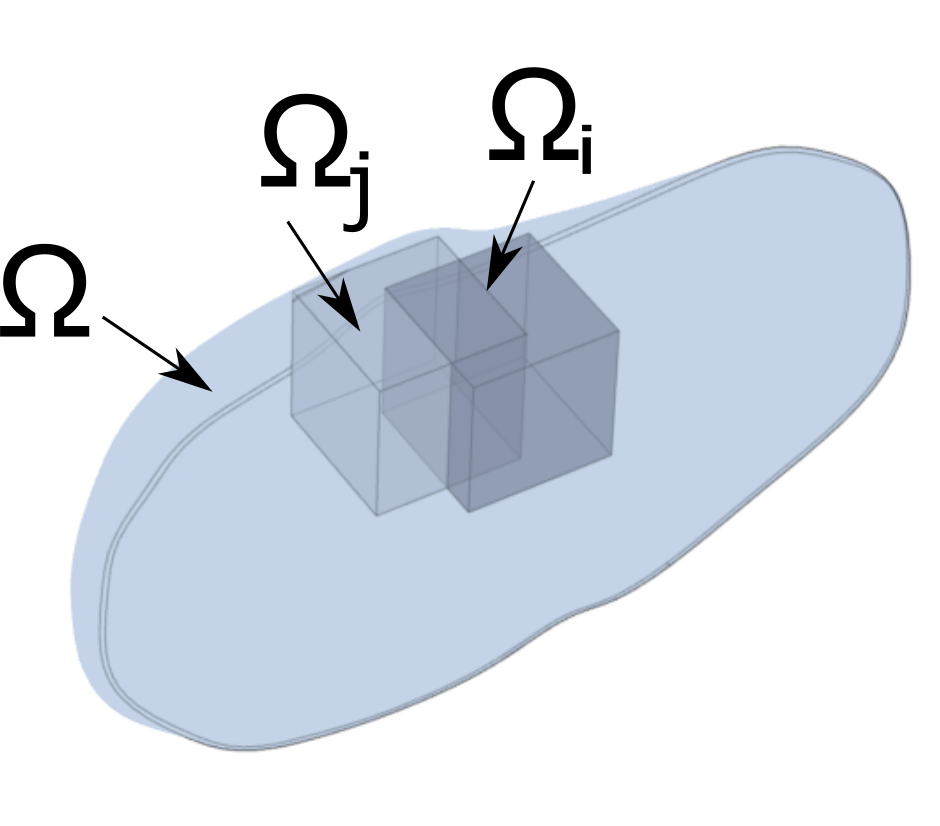}}	
	\caption{A sample of overlapping sub-zones concept showing two sub-zones $\Omega_i$ and $\Omega_j$, where the wave-constraints are solved independently, while the TV prior works on the total problem domain $\Omega$}.
	\label{sub-zones}
\end{figure}

\TMIRevision{For ease of computation, we reformulate (\ref{Optimization Model_MF}) to incorporate the overlapping sub-zones~\cite{VanHouten1999} with a global regularization prior on the elasticity map~\cite{Boyd2010}. We define $\nu_i$, $Q_i$, and $\mathbf{W}_i$ to be the restrictions of  $\mu$, $P$, and $\mathbf{U}$ to local sub-zone~$i$. The sub-zone problem solves for $\left({\mu}^*, {\mathbf{W}}^*, Q^*\right)$ by minimizing:
\begin{align}
\label{Optimization Model_MF2_SZ_Final}
\underset{\begin{subarray}{c}
	\mathbf{\mu} \in \mathbb{C}^{N} \\
	\mathbf{W}_i \in \mathbb{C}^{3MJ}\\
	\nu_i \in \mathbb{C}^{M}\\
	Q_i \in \mathbb{C}^{MJ}\\
	\end{subarray}}{\min} &
 \,\,\,
\Bigg\{\gamma_{\mu} \norm{\mu}_{TV}+ \sum_{i=1}^{N_w}
\frac{\rho}{2} \norm{ \mathbf{W}_i-\mathbb{S}_i\mathbf{V}}^2 \nonumber\\[-20pt]
& \hspace{0.5in} + \sum_{i=1}^{N_w}\left(\gamma_u\norm{\hat{\mathbf{W}}_i}_1+\frac{\gamma_p}{2}\norm{\nabla Q_i}^2\right)
 \nonumber\\
&\Bigg|
\begin{array}{c}
\left[ \mathbb{K}_\mu \left( \nu_i\right)- \mathbb{M}\right]\mathbf{W}_i+ \mathbb{K}_p Q_i= \mathbf{0}  \,\,\, \text{in  } \Omega_i  \\
\nu_i=\mathbb{T}_i\mathbf{\mu} \mbox { and }
\hat{\mathbf{W}}_i=FFT\left({\mathbf{W}_i}\right)
\end{array}
 \Bigg\} 
\end{align}}
\TMIRevision{Here, $\mathbb{S}_i$ is a $3MJ \times 3NJ$ matrix, and $\mathbb{T}_i$ is an $M \times N$ matrix that selects the multifrequency displacements, and the elasticity values for the $i^{th}$ sub-zone, respectively, and $N_w$ is the number of sub-zones. The cost function is minimizing the sum of the cost functions for all the sub-zones. \TMISecondRevision{Each sub-zone follows a wave-constraint model in their respective domain, while the TV regularization prior works on the global elasticity map $\mu$ to reduce the blocking artifact~(Fig.~\ref{sub-zones})}. We have used regularized consensus ADMM~\cite{Boyd2010}.}
\begin{comment}
This allow us to re-formulate (\ref{Optimization Model_MF2}) as a functional to minimize the sum of cost over all the sub-zones. 
\begin{align}
\label{Optimization Model_MF2_SZ}
{\text{minimize   }} & \gamma_{\mu} \norm{\mu_z}_{TV}+ \sum_{i=1}^{N_w}\left(\frac{1}{2} \norm{ \mathbf{U}_i-\mathbf{V}_i}^2+ \gamma_{u} \norm{\mathbf{U}_z,_i}_1\right)\\
\text{subject to}&\left[ \mathbb{K} \left( \mu_i\right)- \mathbb{M}\right] \mathbf{U}_i^{MF}= \mathbf{0} \nonumber\\
& \mathbb{W}\left(\mathbf{U}_i\right)= \mathbf{U}_z,_i \nonumber\\
& \mathbf{\mu_i}= \mathbf{\mu_z,_i}, \hspace{0.2in}   i=1,\cdots, N_w  
\end{align}
Similar to non-linear inversion (NLI), we have divided the volume into overlapping sub-zones, that reduce the computational complexity of the closed-form solution in Elasticity reconstruction and displacement reconstruction steps. To this end, we have formulated the problem as a general form consensus with total variation regularization on the global elasticity map. 
Here all the sub-problem except the elasticity regularization works on each sub-zone separately. The elasticity regularization step involves a local averaging steps of all the different $\mu_i$ and the $z_mu,_i$ to generate a global map, which is subsequently denoised using TV denoising. Using the regularized sub-zone based method avoids the blocking artifact as well as the boundary artifacts occurring in the boundary of each sub-zone.
\end{comment}
%\subsection{Optimization}
\label{Optimization}
\TMIRevision{The augmented Lagrangian function can be written as shown in (\ref{Optimization Model_AL}), where $\alpha_c$, $\alpha_{\mu}$, and $\alpha_W$ are the penalty parameters and $\lambda_{c,i}$, $\lambda_{\mu,i}$, and $\lambda_{W,i}$ are the Lagrange multipliers for the respective equality constraint for the $i^{th}$ sub-zone. The consensus ADMM decomposed (\ref{Optimization Model_AL}) into five sub-problems, which are alternatively solved at each iteration.} \TMISecondRevision{These sub-problems are described in detail in the appendix.} \TMISecondRevision{ In summary, in each iteration we first obtain an estimate of the local elasticity for each sub-zone using the current estimate of displacement. Then, we combine all the local elasticity using the TV prior to obtain the global elasticity. Next, for each sub-zone, new estimates of the local displacement is obtained by fitting the wave equation with the current estimate of shear modulus.  The current estimate of the displacement then denoised using the k-space sparsity. Lastly, dual variables are updated to accumulate the errors in all three equality constraints given in Eq. (\ref{Optimization Model_MF2_SZ_Final}). }\par
\begin{figure*}[!htp]
\begin{align}
\label{Optimization Model_AL}
	\mathcal{L}_\alpha\left(\mu, \nu, \mathbf{W}, Q \right) 
	=&\gamma_{\mu} \norm{\mu}_{TV} + \sum_{\emph{i}=1}^{N_w}\left(\frac{\rho}{2} \norm{ \mathbf{W}_i-\mathbb{S}_i\mathbf{V}}^2+ \gamma_u\norm{\hat{\mathbf{W}}_i}_1+\frac{\alpha_c}{2} \norm{\left[ \mathbb{K}_\mu \left( \nu_i\right)- \mathbb{M}\right]\mathbf{W}_i+ \mathbb{K}_pQ_i+\lambda_c}^2 \right) +  \nonumber \\
	& \hspace{0in} \sum_{i=1}^{N_w} \left(\frac{\gamma_p}{2}\norm{\nabla Q_i}^2+\frac{\alpha_\mu}{2} \norm{\nu_i- -\mathbb{T}_i\,\mu+ \lambda_\mu,_i}^2+\frac{\alpha_W}{2} \norm{FFT\left(\mathbf{W}_i\right)-\hat{\mathbf{W}_i}+ \lambda_{W,i}}^2 \right) 
\end{align}
\end{figure*}
\begin{algorithm}
\label{algo::MERSA}
 \caption{MERSA}
 \begin{algorithmic}[1]
 \renewcommand{\algorithmicrequire}{\textbf{Input:}}
 \renewcommand{\algorithmicensure}{\textbf{Output:}}
 \REQUIRE Measured displacement $\mathbf{V}$ 
 \ENSURE  $\mathbf{U}$, $P$, $\mu$
\STATE Divide the displacement volume into overlapping sub-zones and construct the corresponding mapping matrix $\mathbb{S}_1$,...,  $\mathbb{S}_{Nw}$, and $\mathbb{T}_1$,...,  $\mathbb{T}_{Nw}$
\STATE For all sub-zones: initialize all Lagrange multipliers as the all-zero vector, $\mathbf{W}_i=\mathbb{S}_i\mathbf{V}$,  and $k=1$
\REPEAT
  \FORALL{sub-zones \emph{i}}    
  \STATE construct the matrices $\mathbb{K}_{U}$ and $\mathbb{K}_{P}$   
  \STATE update $\nu_i$ and $Q_i$ with direct inversion using (\ref{Local Elasticity Update})
  \STATE construct the matrices $\mathbb{K} _\mu$ and $\mathbb{M}$ 
  \STATE update $\mathbf{W}_i$ with forward solution using (\ref{Local Displacement Update})
  \STATE update $\hat{\mathbf{W}}_i$ with soft-thresholding (\ref{Displacement_Denoise})
  \ENDFOR
  \STATE update $\mu$ with TV denoising using (\ref{Elasticity_Denoise})
  \FORALL{sub-zones(i)}    
  \STATE update $\lambda_{c,i}, \lambda_{U,i}, \lambda_{Y,i}, \lambda_{\mu,i}$ with (\ref{Dual_Variable_Update4}) %(\ref{Dual_Variable_Update1})-(\ref{Dual_Variable_Update4})
  \ENDFOR
  \STATE k=k+1
\UNTIL $\norm{\mu^k-\mu^{k-1}}_1\leq \text{Tol}_{\mu}$ or $k\geq MaxIter$
 \RETURN $\mu$, $\mathbf{U}$, $P$ 
 \end{algorithmic} 
 \end{algorithm}
% \vspace{-0.2in}
%\subsection{Convergence}
Even though the wave constraint FEM model is bi-convex, global convergence is not guaranteed~\cite{Tan2017,Otesteanu2018d}. However, there is evidence that ADMM provides a faster convergence~\cite{Boyd2010, Mohammed2019}. \TMISecondRevision{In the preliminary version of this work, we have shown empirically for 2D FEM models that ADMM provides better convergence performance than the gradient-based iterative methods~\cite{Mohammed2019}}.       
\section{Method Evaluation}
\label{sec :: Method Evaluation}
Below first we describe the \emph{in silico}, phantom and \emph{in vivo} liver dataset used for the validation of the proposed methods. Then, we present the implementation details for ERSA and MERSA followed by the details of the state-of-the art methods used for comparison and the evaluation metrics.
\subsection{Dataset}
\subsubsection{Numerical Simulations}
\label{subsec:: Numerical Simulations}
\begin{figure}[!t]
	\centering
	{\includegraphics[width=1\linewidth]{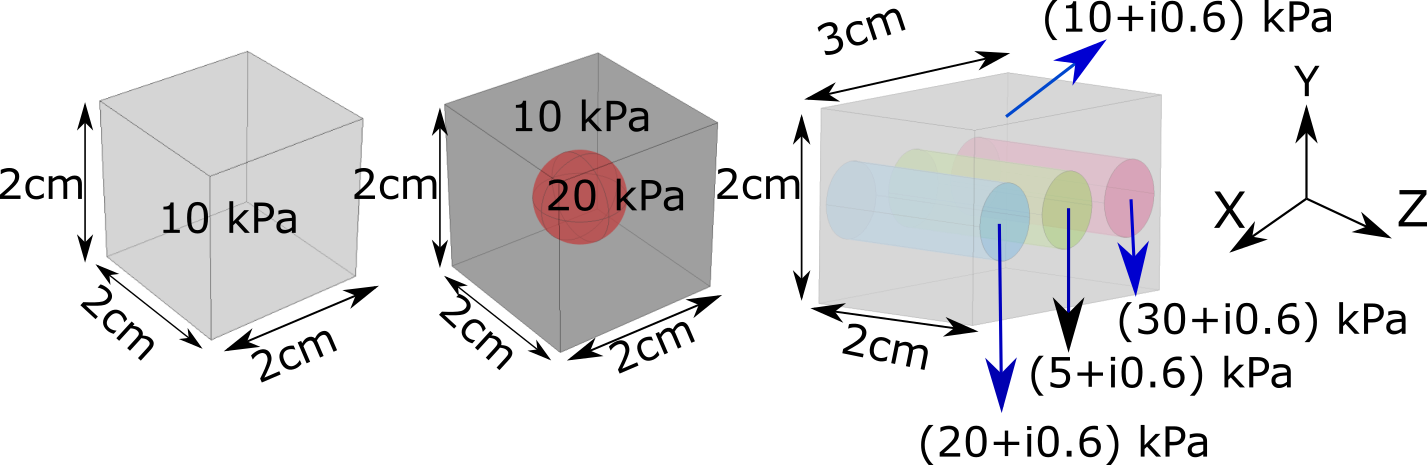}}	
	\caption{\TMIRevision{Numerical models used for generating the 3D synthetic data. For all the models, a harmonic displacement is applied to the bottom surface. The top, front, and the left surfaces are fixed. All the other surfaces are free. The phantoms are (a) Homogeneous  with a storage modulus of 10~kPa and no loss modulus, (b) background storage modulus of 10~kPa, and a spherical inclusion with a storage modulus of 20~kPa, whose radius is varied from 1~mm to 7~mm, and (c) a region with three cylindrical inclusions with different storage moduli. The last phantom has a uniform loss modulus of 0.6~kPa.} }
	\label{Simulation_Phantoms}
\end{figure}

\TMIRevision{Elastic wave simulations using the viscoelastic wave equation given in (\ref{Linearized Equation_Mixed}) were carried out for three models, as shown in Fig. \ref{Simulation_Phantoms}. Here we reuse two models from a previous work, which have been investigated to compare different direct inversion methods~\cite{Honarvar2017comparison}. We use the full harmonic option in ANSYS 2020 R2.  All models are discretized using 10-node tetrahedral elements (solid187) with an element size of 0.5~mm. All three phantoms have a background elasticity of 10~kPA, Poisson's ratio of 0.495, and a density of 1000~kgm$^{-3}$, and are subjected to harmonic frequencies between 10~Hz and 400~Hz.  Their properties are listed in the caption of Fig.~\ref{Simulation_Phantoms}.}
%The first phantom is a homogeneous phantom, for which the displacement pattern is generated for vibration frequencies of 10~Hz to 400~Hz. The second phantom has a  spherical inclusion of 20~kPA placed in the center, whose radius is varied from 1~mm to 7~mm. The displacement pattern is simulated at 200~Hz and 300~Hz vibrations. The viscosity is neglected in these two models.} 
%
\TMIRevision{For the last model, viscosity is incorporated in the wave equation using complex shear moduli. 
% To this end, in ANSYS tabular inputs for storage modulus and loss ratios are defined using the "TB, ELASTIC" and the "TB, SDAMP" options, respectively. 
This model has three cylindrical inclusions of 4~mm radius with elasticity values of 5~kPa, 20~kPa, and 30~kPa. The loss ratio of different regions is chosen to achieve a uniform loss modulus of 600~Pa.} \TMISecondRevision{The simulated displacement from the ANSYS generated mesh is interpolated to a regular grid size of 1.5~mm$\times$ 1.5~mm $\times$ 1.5~mm. 
%The noisy displacement data is generated by adding Gaussian noise to achieve a signal to noise ratio (SNR) of 25~dB.
} \par
%\vspace{-0.1in}
\subsubsection{Phantom Experiments}
\label{subsec:: Phantom Experiment}
 The method was tested on two set of phantoms. A cohort of liver mimicking elasticity phantoms (Model 039, CIRS Inc., Norkflok, USA ) and a quality assurance phantom (Model 049, CIRS Inc., Norfolk, USA) with four different spherical inclusions. The phantoms' characteristics are described in the results section. The CIRS 039 model was imaged at a voxel size of 1.705~mm$\times$ 1.705~mm at four frequencies of 50, 55, 60 and 65~Hz with a gradient echo MRE sequence \cite{Sahebjavaher2013a}. \TMIRevision{The slice thickness was 5~mm and the number of slices is 16}. \TMIRevision{Compared to the CIRS039 phantoms, the CIRS 049 phantom has a higher elasticity. As a result, we have used a higher vibration frequency of 210~Hz and 250~Hz with a spin-echo MRE sequence (\cite{Mohammed2019}) at a voxel size of 1.38~mm $\times$ 1.38~mm and at a slice thickness of 1.5~mm}. The mechanical motion was captured at 8 states for the CIRS049 and 4 states for the CIRS039 phantom.  All experiments were performed on a Philips Achieva 3T system (Philips Inc., Netherlands).\par
 
\TMIRevision{A CIRS-provided certificate was used as a gold standard for the CIRS049 phantom. 
For the CIRS039 phantoms, we report both the manufacturer's range (this has been revised by CIRS since the purchase of the phantom), and measured reference elasticity values using a Verasonics Vantage\texttrademark~scanner following the QIBA protocol \cite{QIBA:2020}. 
We did not use the QIBA protocol on the CIRS049 phantom due to the presence of heterogeneity. }
% To differentiate between the initially reported elasticity and the adjusted values, we have termed these values as QIBA. {\bf Do we have the initially reported values in this paper?} }}\par  
%\vspace{-0.1in}
\subsubsection{ \emph{In vivo} Experiment}
\label{subsec:: In-vivo Experiment}
After ethics approval from the University of British Columbia Clinical Research Ethics Board, the proposed method was tested in liver MRE in five healthy volunteers who have provided informed consent. \TMIRevision{All the scanning was performed using the Philips Elition 3T X scanner}. \TMISecondRevision{A breath-hold MRE sequence was used to image the displacements at four vibration frequencies 50, 55, 60, and 65~Hz. A four point unbalanced encoding strategy that imaged the displacement in three orthogonal directions: slice, measurement and phase, and a reference scan with disabled motion encoding gradient was used for capturing the three-directional MRE.} \TMIRevision{The reconstruction voxel size for the displacement was 1.7~mm$\times$ 1.7~mm with a slice thickness of 5~mm. Each frequency and direction of the displacement consisted of eight slices and four dynamics.} \TMIRevision{Note that only one displacement direction and frequency was captured per breath-hold.} Each breath-hold was less than 18 seconds. \TMIRevision{No registration was applied between the displacements from different frequencies and directions}. A T2W sequence was used with the same field of view as the MRE to generate the anatomical reference of the liver. In addition, 2D MRE with the Resoundant system was also performed with the following settings: resolution: 1.17~mm $\times$ 1.17~mm; slice thickness: 10~mm; slice gap: 1~mm; vibration frequency: 60~Hz. \TMISecondRevision{The reconstructed elastogram and associated confidence map for Resoundant was obtained from the scanner.}  
\vspace{-0.1in}
\subsection{Implementation Details for ERSA and MERSA}
\label{subsec:: Preprocessing}
\label{subsec:: Implementation}
\TMIRevision{The displacement fields captured for both phantom and \emph{in vivo} data are first transformed to displacement phasors by taking the Fourier transform in time and taking the fundamental frequency component.
The displacement phasors are sequentially phase unwrapped and interpolated to an isotropic grid. For all cases, we interpolate the slices in  the in-plane resolution. A three-dimensional plane fitting method was used to calculate the numerical derivative with a span of 3 pixels for finding the displacement gradient for elasticity inversion in both MERSA and Mixed-FEM. }\par 
%\vspace{-0.1in}
%\subsection{Implementation Details}
\TMIRevision{ 
All the regularization parameters were optimized empirically from a single numerical phantom dataset. For this reason, we have used the heterogeneous numerical phantom with the 5~mm spherical inclusion as shown in Fig.~\ref{Simulation_Phantoms}. The simulated displacement from the phantom is contaminated with Gaussian noise to achieve a SNR of 25~dB. The regularization parameters were optimized to achieve the lowest error. For all the other experiments, the same set of parameter values was used without fine tuning. 
Sub-zones of 21~mm$\times$21~mm$\times$21~mm were used with a stride of 17~mm in each dimension. 
The initial guess for the shear modulus $\mu$ was 3~kPa+0i~kPa for all the experiments, while the measured displacement is taken as the initial guess for the fitted displacement (i.e., $\mathbf{W}_i^0= S_i \mathbf{V}$). 
We use an FEM system with an eight-node quadratic shape function for displacement and constant shape functions for hydrostatic pressure and shear modulus. 
} \par

\TMIRevision{
The regularization parameters, along with their normalizing factors, are listed in Table \ref{table_reg_param}.
The regularization parameter $\alpha_{\mu}$ was selected to obtain an estimated condition number for the inversion $\left(\alpha_c\mathbb{K}_{u}^T\mathbb{K}_{u}+\alpha_{\mu} \mathbb{I}\right)$ equal to $2^{12}$. 
In this case, to calculate $\mathbb{K}_{u}$, the measured displacement $V$ is used. 
Similarly, the value of $\rho$ is selected to obtain an estimated condition number for the inversion $\left(\alpha_c \left[ \mathbb{K} - \mathbb{M}\right]^T\left[ \mathbb{K}- \mathbb{M}\right]+ \rho\mathbb{I}\right)$ equal to $2^4$. 
In this regard, we use the shear modulus $\mu^{1}$, which is the estimate after the first global elasticity update.} \par 
\TMIRevision{Instead of using an absolute value for the regularization parameter for priors ($\gamma_\mu,\gamma_u$), we selected relative values with respect to the maximum value of the projection in the respective sparsity transform \cite{Boyd2010}. 
% We have selected  $\gamma_\mu$ empirically to be $2^{-14}$ of the maximum value of the TV projection of the shear modulus for the TV prior. 
% For all cases, we used $\alpha_X= 10^{-3} \rho$, and $\gamma_u= 2^{-7}$. 
%{\bf <- Perhaps we can eliminate repeats from Table I?}
}
\begin{table}[!t]
\caption{Regularization parameters used in all the experiments for ERSA and MERSA. Here, $MaxEig$ and $MaxAbs$ calculates the maximum Eigenvalue and Maximum absolute values respectively. }
\label{table_reg_param}
\centering
%% Some packages, such as MDW tools, offer better commands for making tables
%% than the plain LaTeX2e tabular which is used here.
\begin{tabular}{|c|c|}
\hline
Parameter & Value \\
\hline
$\alpha_c$ & $1$\\
\hline
$\alpha_\mu$ & $\alpha_c\times MaxEig\left( \mathbb{K}_{u}^T\mathbb{K}_{u}\right) \times 2^{-12}$\\
\hline
$\rho$ &$MaxEig\left( \left[ \mathbb{K}_\mu- \mathbb{M}\right]^T  \left[ \mathbb{K}_\mu- \mathbb{M}\right]\right)\times2^{-4}$\\
\hline
$\alpha_X$ &$10^{-3} \rho$ \\
\hline
$\alpha_W$ &$MaxEig\left( \left[ \mathbb{K}_\mu- \mathbb{M}\right]^T  \left[ \mathbb{K}_\mu- \mathbb{M}\right]\right)\times2^{-4}\times10^{-2}$\\
\hline
$\gamma_u$ &$MaxAbs\left( FFT\left(v\right) \right)\times2^{-7}$\\
\hline
$\gamma_\mu$ &$MaxAbs\left( \nabla \mu \right)\times2^{-14}$\\
\hline
$\gamma_p$ &$\frac{MaxEig\left( \mathbb{K}_{p}^T\mathbb{K}_{p} \right)}{MaxEig\left({\nabla}^T{\nabla}\right)}\times2^{-16}$\\
\hline
$Tol_\mu$ & $10^{-3}$\\
\hline
MaxIter & $100$\\
\hline

\end{tabular}
\end{table}
\vspace{-0.1in}
\subsection{Performance Comparison with Other Methods}
\label{subsec:: Comparison}
\TMIRevision{To evaluate the performance of the proposed method, we compare the results with direct inversion methods such as sparsity regularized based mixed FEM (Mixed-FEM)~\cite{Honarvar2012}, local frequency estimation (LFE)~\cite{Manduca1996}, and tomoelastography (kMDEV)~\cite{Tzschatzsch2016}. 
For Mixed-FEM, we utilize the implementation detailed in \cite{Honarvar2016}, which uses the sparsity regularization for better conditioning of (\ref{Linearized Equation_IP}). 
The sparsity regularization is achieved by applying a truncated Discrete Cosine Transform as a pre-multiplier to the unknown parameter. 
In turn, this reduces the rank of the inverse problem and provides robustness to noise. 
Since Mixed-FEM is a single frequency implementation, for the multifrequency results shown for CIRS049 and \emph{in vivo} liver volunteer data, we take the average of the single frequency implementation. }\par

\TMIRevision{In addition to Mixed-FEM, we also use two well-known reconstruction algorithms: local frequency estimation (LFE)~\cite{Manduca1996} and kMDEV \cite{Tzschatzsch2016}. 
LFE assumes the measured displacement to be due to a shear wave and uses spatial filter banks to estimate the local wavelength from the measured displacement~\cite{Manduca1996}. 
This work used the LFE implementation given in~\cite{baghani2011travelling, Honarvar2017comparison}. 
We used a fourth order Butterworth bandpass filtering with a range from 1~kPa-40~kPa for synthetic phantoms, CIRS039 phantoms, and the \emph{in vivo} liver data. 
For the CIRS049 phantom, the bandpass filtering range was changed to 1~kPa-70~kPa. 
kMDEV is a 2D method that decomposes the measured displacement into plane waves moving in different directions and estimates the local wavelength from each direction with the finite difference method \cite{Tzschatzsch2016}.
We have used the implementation available online at BIOQIC~\cite{KMDEV:2009}. 
The default settings were used for CIRS039 phantoms and the \emph{in vivo} liver data. We have changed the smoothing strength to 0.1~mm from 2.75~mm for the CIRS049 phantom with feedback from BIOQIC. For the numerical phantoms, the smoothing strength of 0.7~mm is used, which corresponds to the best performance in terms of RMSE for the single inclusion numerical phantom with an inclusion radius of 5~mm. 
Since kMDEV produces shear speed $c$ and attenuation $a$, the complex shear modulus is calculated by:
\begin{align}
\mu= \frac{\rho}{\left(\frac{1}{c}-i\frac{1}{2\pi a}\right)^2} \,\,.
\end{align}
However, as the attenuation $a$ is affected more by noise than $c$, we used the simplified $3\rho c^2$ for the storage modulus. Both LFE and kMDEV support multi-direction and multifrequency reconstruction using weighted averaging of the single frequency reconstruction.}\par

\TMIRevision{We evaluate the usefulness of using the multifrequency reconstruction method by comparing the performance of MERSA with multifrequency displacement and single frequency displacement. For ERSA, we have used the same regularization parameter as given in Table \ref{table_reg_param}.}\par 
\TMIRevision{ For the CIRS039 phantom and 
\emph{in vivo} liver study, we compare the results with elasticity measured by the Resoundant system. 
However, the vibration frequency of 60Hz was not sufficiently high for the CIRS049 phantom and these results are not reported.}
\begin{comment}
The DIFEM were carried out with a least square solver with Tikhonov regularization\cite{Park2006}. For both  DIFEM and LFE, a bandpass filter was applied as a preprocessing stage to remove high frequency measurement noise and low frequency compressional wave from the measured displacement. Filter bandwidth and the regularization parameter were selected in order to maximize the peak signal to noise ratio (PSNR). \par

The iterative FEM(IterFEM) method utilizes a gradient based method, where the gradient are calculated using a adjoint based method \cite{Tan2017}. Both the original IterFEM \cite{VanHouten1999} and hybrid FEM (HFEM)\cite{Otesteanu2018d} used a 2D single frequency implementation, we have used the 3D multifrequency implementation in this paper. In addition, 

\end{comment}
\subsection{Evaluation Metrics}
\label{subsec:: EvaluationMetrics}
\TMIRevision{We evaluated the reconstructed elasticity using the Root-mean-squared-error (RMSE) defined by: 
\begin{align}
    RMSE= \sqrt{\frac{1}{m} \norm{\frac{\mu_{rec}-\mu_{GT}}{\mu_{GT}}}_1}
\end{align}
where $\mu_{rec}$ and $\mu_{GT}$ are the reconstructed elasticity and the ground truth elasticity respectively. In case we are showing the results for elasticity and viscosity separately, we use $\text{RMSE}_\text{E}$ and $\text{RMSE}_\text{V}$ to show the error for elasticity and viscosity respectively.}\par
\TMIRevision{For elasticity we have also used the contrast-to-noise ratio (CNR), which is a measure of the sharpness of reconstruction between two different homogeneous regions denoted as background and inclusion. For example, if the mean and standard deviation of elasticity for background and inclusion are $\mu_{bkg}$, $\mu_{inc}$, $\sigma_{bkg}$ and $\sigma_{inc}$ respectively, then the CNR is given by:
\begin{align}
    CNR=\frac{2\left(\mu_{inc}-\mu_{bkg}\right)^2}{\sigma_{bkg}^2+\sigma_{inc}^2}
\end{align}
For elasticity reconstruction of the CIRS phantom, we have measured the two-way interclass correlation (ICC) with the reported elasticity values~\cite{Solamen2018,ICC_code}}. 
\vspace{-0.1in}
\section{Results}
\label{sec :: Results}
This section describes various experiments conducted to quantitatively and qualitatively evaluate the performance of MERSA.
Implementation settings are summarized in \ref{subsec:: Implementation}.
To avoid duplication, specific experimental details are provided only in the figure captions.
\vspace{0in}
\subsection{Effect of Mesh Size to Wavelength Ratio on the Accuracy of Elasticity Reconstruction}
\label{subsec: In-Silico-Homo}
\begin{figure}[!t]
\centering
{\includegraphics[clip,width=0.8\linewidth]{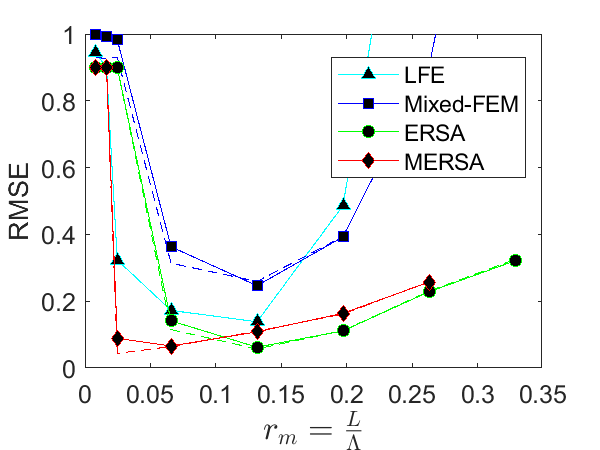}}
\caption{\TMISecondRevision{ Elasticity RMSE versus $r_m$, the ratio of voxel size to wavelength, for single frequency Mixed-FEM and ERSA and multifrequency LFE and MERSA using the {\em in silico} homogeneous phantom. The frequencies are 10~Hz, 20~Hz, 30~Hz, 80~Hz, 160~Hz, 320~Hz and 400~Hz. The multifrequency implementations (MERSA and LFE) use these seven sets of frequencies: (10,20), (20,30), (30,80), (80,160), (160,240), (240,320), (320, 400). For the multifrequency implementation, we calculated the $r_m$ using the lower frequency.  For each colour, the dashed line shows reconstructions with noiseless data.}}
\label{Homo_Plot2}
\end{figure}

\TMIRevision{The homogeneous numerical phantom's elasticity was reconstructed for both noiseless and noisy displacements using ERSA, MERSA, LFE, and Mixed-FEM.  
The results of the RMSE with voxel ratio to wavelength ($r_m$) are shown in Fig. \ref{Homo_Plot2}.
\TMISecondRevision{ERSA, LFE, and MERSA are less affected by $r_m$ compared to mixed-FEM. ERSA performs better than MERSA at $r_m\geq 0.13$}.} 

\begin{comment}
\begin{figure*}[!t]
\centering
\subfloat[Noiseless]{\includegraphics[clip,width=0.4\linewidth]{./Figures/Experiment1/All_Homoplot.eps}} 
\hspace{0.05\linewidth}
\subfloat[10 dB SNR]{\includegraphics[clip,width=0.4\linewidth]{./Figures/Experiment1/All_Homoplot.png}}
\caption{ Summary results from homogeneous synthetic data shows the mean estimated stiffness vs. the original elasticity of the region for MERSA and multifrequency LFE for two cases: (a) Noiseless and (b)10 dB SNR. The corresponding standard deviation is shown as an error bar at each point. }
\label{Homo_Results}
\end{figure*}
\end{comment}
\vspace{-0.1in}
\subsection{Detectability of Inclusions}
\label{subsec: In-Silico-Hetero}
\begin{figure}[!t]
\centering
{\includegraphics[clip,width=0.9\linewidth]{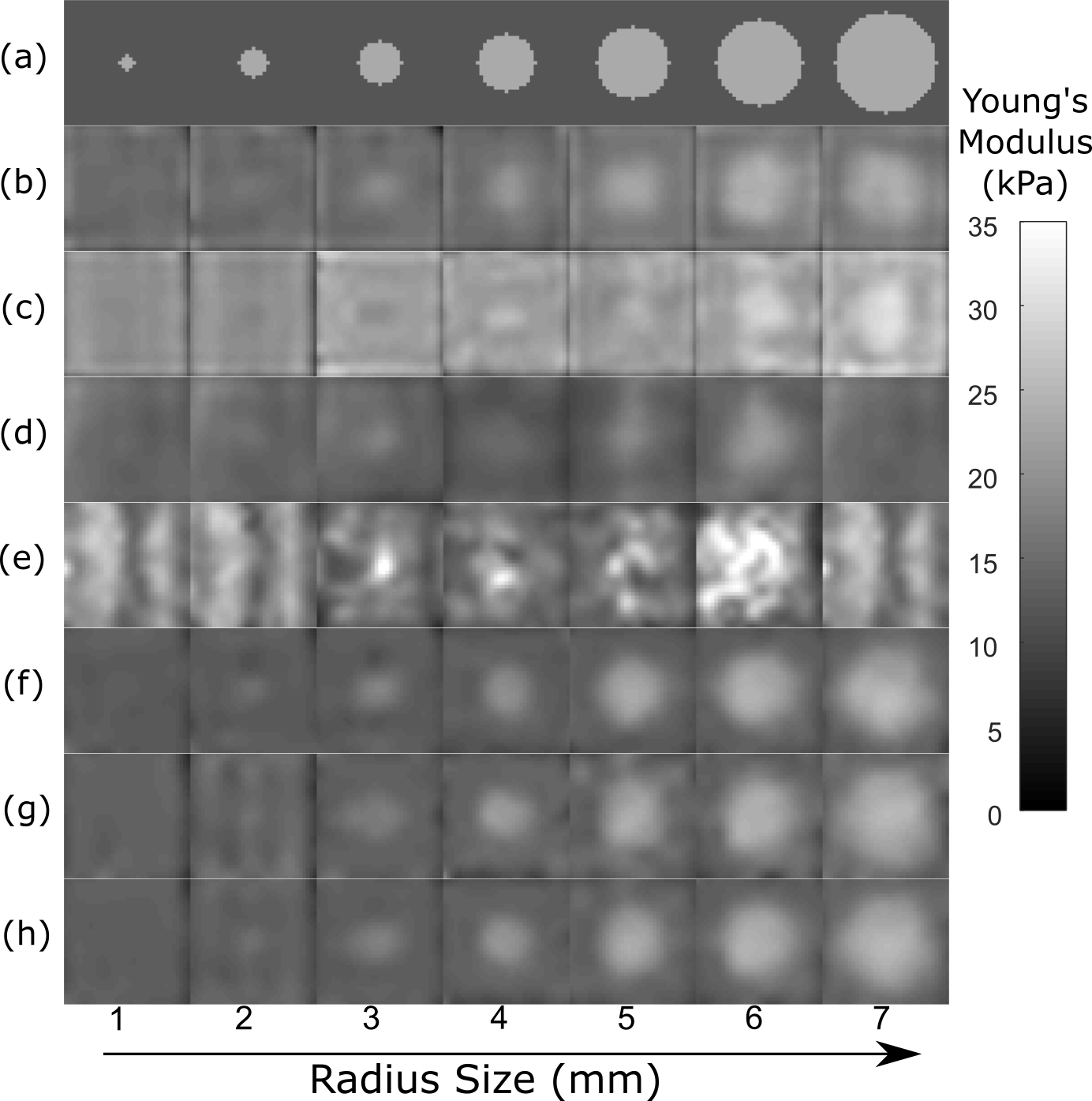}}
\caption{\TMISecondRevision{Elastograms for {\em in silico} inclusions with varying size with noisy displacements. (a) Ground truth, (b) Mixed-FEM at 200~Hz, (c) Mixed-FEM at 300~Hz, (d) LFE with 200~Hz and 300~Hz, (e) kMDEV with 200~Hz and 300~Hz, (f) ERSA at 200~Hz, (g) ERSA at 300~Hz, and (h) MERSA with 200~Hz and 300~Hz. The intra-region variance due to noisy displacements is most pronounced for Mixed-FEM. 
ERSA ((d)-(e)) and (f) MERSA reduce the effect of noise in both the background and inclusion.}}
\label{Hetero1Inc_Image}
\end{figure}

\begin{figure}[!t]
\centering
{\includegraphics[clip,width=0.8\linewidth]{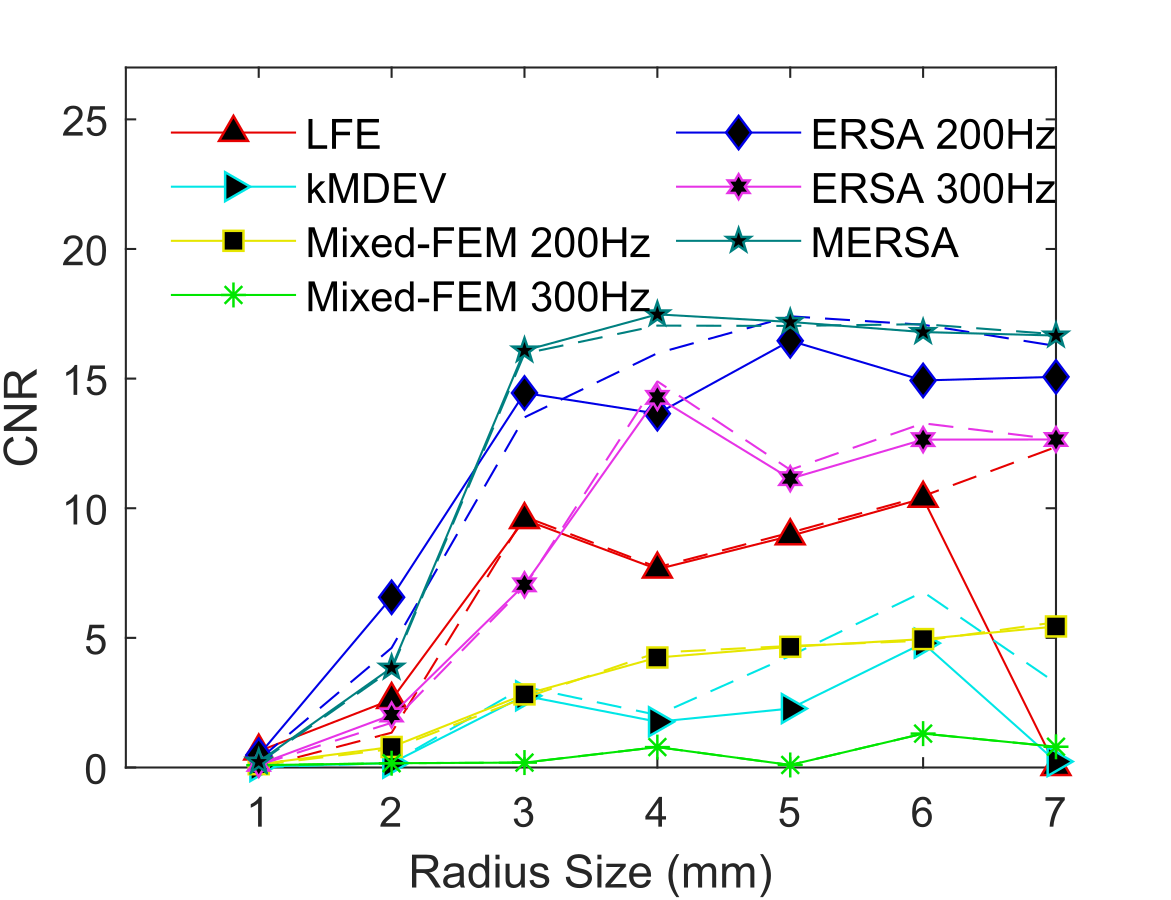}}
\caption{\TMISecondRevision{Comparison of CNR with inclusion size for Mixed-FEM, LFE, ERSA, and MERSA in noiseless (dash line) and noisy case (solid line) for the {\em in silico} single inclusion phantom.
%{\bf $\leftarrow$ I would prefer this plots to be provided in a single plot, coloured with different tic marks (circles, x-es, crosses, etc), so they can be seen on the same plot }
}}
\label{Hetero1Inc_Plot}
\end{figure}

\TMIRevision{The reconstruction results for inclusions of radii from  1~mm to 7~mm are given in Fig. \ref{Hetero1Inc_Image}. 
\TMISecondRevision{The effect of discretization is more prominent for the single frequency Mixed-FEM implementation.} 
ERSA is less affected by discretization and provide lower intra-region variance; MERSA produce the best CNR, as shown in the CNR comparison plot in Fig. \ref{Hetero1Inc_Plot}.}\par

\subsection{Viscoelastic Reconstruction}
\label{subsec: In-Silico-ViscoHetero}
\begin{figure*}[!t]
{\includegraphics[clip,width=1\linewidth]{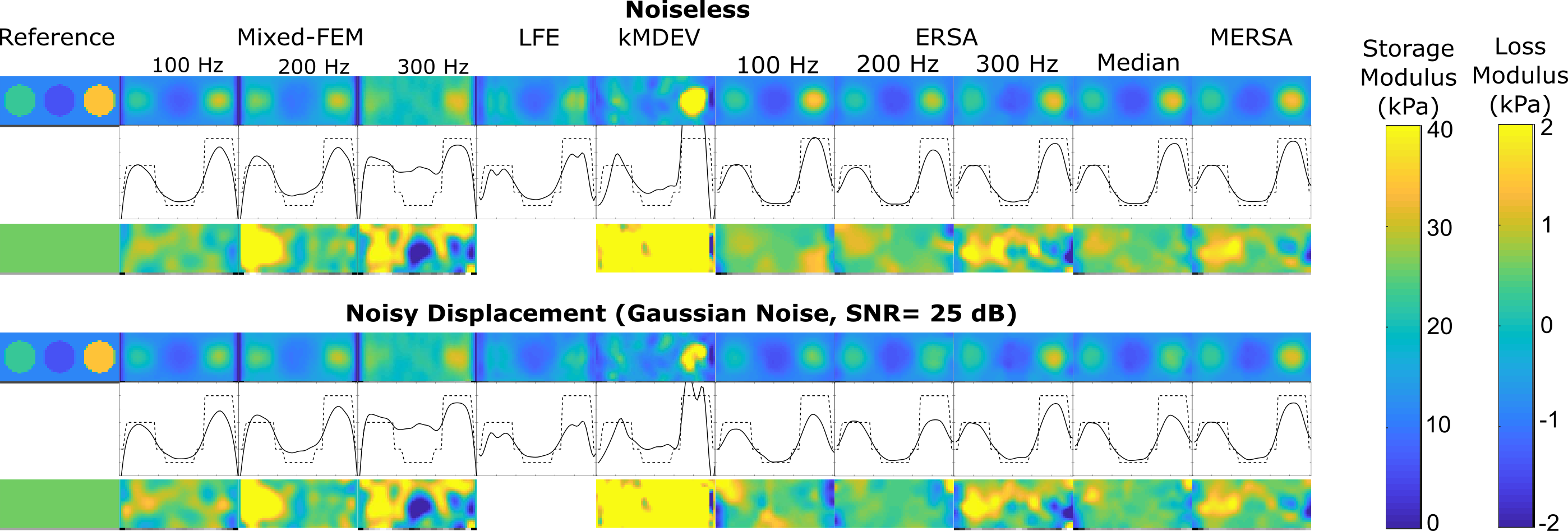}}
\caption{\TMISecondRevision{Comparison of reconstructed storage and loss moduli for the multi-inclusion numerical phantom, computed using noiseless displacement (top three rows) and noisy displacements (bottom three rows). The first and last row of each grouping show the loss modulus and storage modulus, respectively. The middle rows shows the loss modulus variation along the mid-horizontal line. }}
%{\bf $\leftarrow$ Change caption from 25dB Noisy Displacement to "Noisy Displacements (Gaussian Noise, SNR = 25 dB)"} }
\label{Hetero3Inc_Image}
\end{figure*}

\TMIRevision{The reconstruction quality comparison on the multi-inclusion phantom with homogeneous viscosity for LFE,  Mixed-FEM, ERSA, and MERSA is shown in Fig. \ref{Hetero3Inc_Image}, and the statistics for RMSE and CNR are listed in Table \ref{table_Hetero3Inc}. As can be seen, MERSA provides the least RMSE for both elasticity and viscosity for most cases. Also, MERSA provides the best CNR for all the inclusions. However, for ERSA, noise in displacement cause artifact in the stiffer region.} \TMISecondRevision{Taking the median of the ERSA results does not remove these artifacts. These artifacts are less prominent in MERSA due to the joint reconstruction of multifrequency displacement.} \par

\begin{comment}
\begin{table}[!t]
%% increase table row spacing, adjust to taste
\renewcommand{\arraystretch}{1.3}
% if using array.sty, it might be a good idea to tweak the value of
% \extrarowheight as needed to properly center the text within the cells
\caption{\TMIRevision{Comparison of RMSE for loss ($\text{RMSE}_\text{E}$) and storage ($\text{RMSE}_\text{V}$) moduli and CNR for the Multi-Inclusion Numerical Phantom. The best results are shown in bold.}}
\label{table_Hetero3Inc}
\centering
%% Some packages, such as MDW tools, offer better commands for making tables
%% than the plain LaTeX2e tabular which is used here.
\resizebox{\columnwidth}{!}{%
\begin{tabular}{|c|c|c|c|c|c|c|c|c|}
\hline
\multirow{2}{*}{}&\multicolumn{3}{|c|}{Mixed-FEM} & \multirow{2}{*}{LFE}& \multicolumn{3}{|c|}{ERSA} & \multirow{2}{*}{MERSA}\\
\cline{2-4} \cline{6-8}
&100&200&300&&100&200&300&\\
\hline
\hline
$\text{RMSE}_\text{E}$& 0.43 & 0.28 & 0.18 & 0.44& 0.21 & $\textbf{0.13}$& 0.15&$\textbf{0.13}$  \\
\hline
$\text{RMSE}_\text{V}$& 1.91 & 2.12 & 1.31& -&1.24 & 0.95 & 0.61& \textbf{0.41} \\
\hline
$\text{CNR}_\text{ 5}$&3.7 &6.7 & 14.1 & 5.1& 20.4 &48.2 & 61.3 & \textbf{66.2}\\
\hline
$\text{CNR}_\text{20}$ &0.1 & 1.6 & 3.1 & 0.6& 1.4 & 12.3 & \textbf{25.5}& 23.1 \\
\hline
$\text{CNR}_\text{30}$ &0.0 & 0.0 & 5.7 & 2.2& 9.2& 20.2 & 23.2& \textbf{24.0} \\
\hline
\end{tabular}}
\end{table}
\end{comment}

\begin{table}[!t]
%% increase table row spacing, adjust to taste
\label{Tabel:: Hetero3Inc}
\renewcommand{\arraystretch}{1.3}
% if using array.sty, it might be a good idea to tweak the value of
% \extrarowheight as needed to properly center the text within the cells
\caption{\TMISecondRevision{Comparison of RMSE for loss ($\text{RMSE}_\text{E}$) and storage ($\text{RMSE}_\text{V}$) moduli and CNR for the Multi-Inclusion Numerical Phantom. MF represents multifrequency implementation using 100~Hz, 200~Hz and 300~Hz frequencies. The best results are shown in bold.}}
\label{table_Hetero3Inc}
\centering
%% Some packages, such as MDW tools, offer better commands for making tables
%% than the plain LaTeX2e tabular which is used here.
\resizebox{\columnwidth}{!}{%
\begin{tabular}{|c|c||c|c|c|c|c|}
\hline
Method& Frequency&$\text{RMSE}_\text{E}$&$\text{RMSE}_\text{V}$&$\text{CNR}_\text{ 5}$&$\text{CNR}_\text{ 20}$&$\text{CNR}_\text{ 30}$\\
\hline
\multirow{3}{*}{Mixed-FEM}& 100& 0.34 & 1.63 & 32.8 & 32.8 & 35.6\\
& 200& 0.38 & 1.78 & 19.4 &25.3 &31.2\\
& 300& 0.51 & 2.71 & 0.0& 9.1 & 22.1\\
\hline
LFE & MF& 0.34 & -& 10.3& 2.5& 6.3\\
\hline
kMDEV& MF & 0.38 & 7.1 & 1.3& 1.7& 2.5\\
\hline
\multirow{4}{*}{ERSA} & 100 & 0.25 & 1.64 & 33.6 & 10.0 & 13.7\\
& 200 & 0.24 & 1.37 & 36.1 & 23.2 & 21.9\\
& 300 & 0.19 & 1.53 & 32.2 &44.8 & 47.5\\
& Median &0.20 & 0.95 & 53.9&27.9 &43.4\\
\hline
MERSA & MF&\bf{0.15} & \bf{0.86} & \bf{81.9} & \bf{54.2} & \bf{57.5}\\
\hline
\end{tabular}}
\end{table}

%\vspace{-0.2in}
\subsection{Tissue Mimicking Phantoms}
% TS - repeat from methods We reconstructed elastograms for four homogeneous CIRS039 phantoms, representing the liver elasticity at different stages of fibrosis. 
The mean and standard deviation of a region of interest defined by a \TMIRevision{20~mm radius circle was calculated for Mixed-FEM, LFE, kMDEV, Resoundant and MERSA. 
% TS - repeated from methods - We also report elasticity from CIRS039 and reconstructed elasticity from Philips Resoundant MRE. 
% TS - repeat - We have also calculated the phantom's elasticity using the protocol defined by baBA \cite{QIBA:2020} with Verasonics Vantage\texttrademark. 
The results are shown in Table \ref{table_CIRS039Homo}.
MERSA gives elasticity values close to the elasticity measured by QIBA for the first three phantoms. 
However, the elasticity is underestimated by around $15\%$ for the stiffest phantom because of the low $r_m$ values, which resulted in an around $5\%$ lower ICC value compared to Resoundant. Compared to QIBA, we see Resoundant consistently overestimates the elasticity value for all the phantoms except the stiffest phantom. 
%{\bf $\leftarrow$ we do need to address why Resoundant, which also uses dynamic elastography and low frequency, produces higher values.  
%Also the ICC is lower...}
} \par  
\begin{comment}
\begin{table*}[!t]
%% increase table row spacing, adjust to taste
%\renewcommand{\arraystretch}{1.3}
% if using array.sty, it might be a good idea to tweak the value of
% \extrarowheight as needed to properly center the text within the cells
\caption{\TMISecondRevision{Reconstructed elasticity for the CIRS~039 Homogeneous phantoms by various methods, expressed in kPa with mean $\pm$ standard deviation format.}}
\label{table_CIRS039Homo}
\centering
%% Some packages, such as MDW tools, offer better commands for making tables
%% than the plain LaTeX2e tabular which is used here.
\resizebox{\columnwidth}{!}{%
\begin{tabular}{|c|l||l|l|l|l|l|l|}
\hline
CIRS & QIBA & Mixed& LFE & kMDEV & Resoun- & MERSA \\
 &  & -FEM&  &  & dant &  \\
\hline
\hline
3.7 &   $2.5$   &$2.5$      &$2.3$      & $2.8$     &$2.7$      & $2.5$ \\  
    &$\pm0.2$   &$\pm0.1$   &$\pm0.0$   & $\pm0.6$  &$\pm0.1$   & $\pm0.0$\\
\hline
13&$6.3$&$6.5$ &$6.3$& $7.6$  &$7.2$ & $6.7$ \\
&$\pm0.5$&$\pm0.3$ &$\pm0.2$& $\pm1.9$  &$\pm0.3$ & $\pm0.4$ \\
\hline
25&$15.9$&$15.4$ &$14.3$& $17.5$  &$18.1$ & $15.2$\\
&$\pm0.4$&$\pm0.2$ &$\pm1.2$& $\pm4.9$  &$\pm1.3$ & $\pm0.7$\\
\hline
50&$33.3$&$17.8$ &$25.5$& $17.6$ &$32.3$ & $28.4$ \\
&$+1.7$&$\pm3.7$ &$\pm3.0$& $\pm4.8$ &$\pm3.7$ & $\pm2.4$ \\
\hline
ICC
&$0.93$&$0.59$&$0.86$ & $0.56$ & $0.91$ & $0.86$ \\
\hline
\end{tabular}}
\end{table*}
\end{comment}

\begin{table}[!t]
%% increase table row spacing, adjust to taste
%\renewcommand{\arraystretch}{1.3}
% if using array.sty, it might be a good idea to tweak the value of
% \extrarowheight as needed to properly center the text within the cells
\caption{\TMISecondRevision{Reconstructed elasticity for the CIRS~039 Homogeneous phantoms by various methods, expressed in kPa with mean $\pm$ standard deviation format. The last column shows the interclass correlation values between the mean elasticity and the reported elasticity by CIRS. }}
\label{table_CIRS039Homo}
\centering
%% Some packages, such as MDW tools, offer better commands for making tables
%% than the plain LaTeX2e tabular which is used here.
\resizebox{\columnwidth}{!}{%
\begin{tabular}{|c|c|c|c|c|c|}
\hline
 & Phantom 1 & Phantom 2& Phantom 3 & Phantom 4 & ICC \\
\hline
\hline
CIRS & 3.7 & 13 & 25 & 50 & -\\
\hline
QIBA & $2.5\pm0.2$ & $6.3\pm0.5$& $15.9\pm0.4$& $33.3\pm1.7$& $0.93$\\
\hline
Mixed-FEM & $2.5\pm0.1$ & $6.5\pm0.3$ & $15.4\pm0.2$ & $17.8\pm3.7$& $0.59$\\
\hline
LFE & $2.3\pm0.0$& $6.3\pm0.2$ & $14.3\pm1.2$ & $25.5\pm3.0$& $0.86$\\
\hline
kMDEV & $2.8\pm0.6$ & $7.6\pm1.9$ & \TMISecondRevision{$16.9\pm5.2$}&\TMISecondRevision{$22.3\pm10.2$} & \TMISecondRevision{$0.70$} \\
\hline
Resoundant & $2.7\pm0.1$ & $7.2\pm0.3$ & $18.1\pm1.3$ & $32.3\pm3.7$ & $0.91$\\
\hline
\TMISecondRevision{ERSA 50}& $2.4\pm0.1$&$6.5\pm0.1$ & $15.0\pm0.5$&$15.0\pm3.8$ &$0.49$\\
\hline
\TMISecondRevision{ERSA 55}&$2.4\pm0.0$ & \TMISecondRevision{$6.8\pm0.4$}& \TMISecondRevision{$15.5\pm1.3$}&\TMISecondRevision{$24.0\pm2.2$}  &\TMISecondRevision{0.76}\\
\hline
\TMISecondRevision{ERSA 60}&$2.5\pm0.1$ & $6.8\pm0.2$&$12.2\pm1.3$ &$14.0\pm4.2$ &$0.44$\\
\hline
\TMISecondRevision{ERSA 65}& $2.7\pm0.1$& $7.0\pm0.2$&$16.3\pm0.8$ &$24.4\pm6.0$ &$0.77$\\
\hline
MERSA & $2.5\pm0.0$ & $6.7\pm0.4$ & $15.2\pm0.7$ & $28.4\pm2.4$& $0.86$\\
\hline
\hline
\end{tabular}}
\end{table}

\begin{figure}[!t]

{\includegraphics[clip,width=1\linewidth]{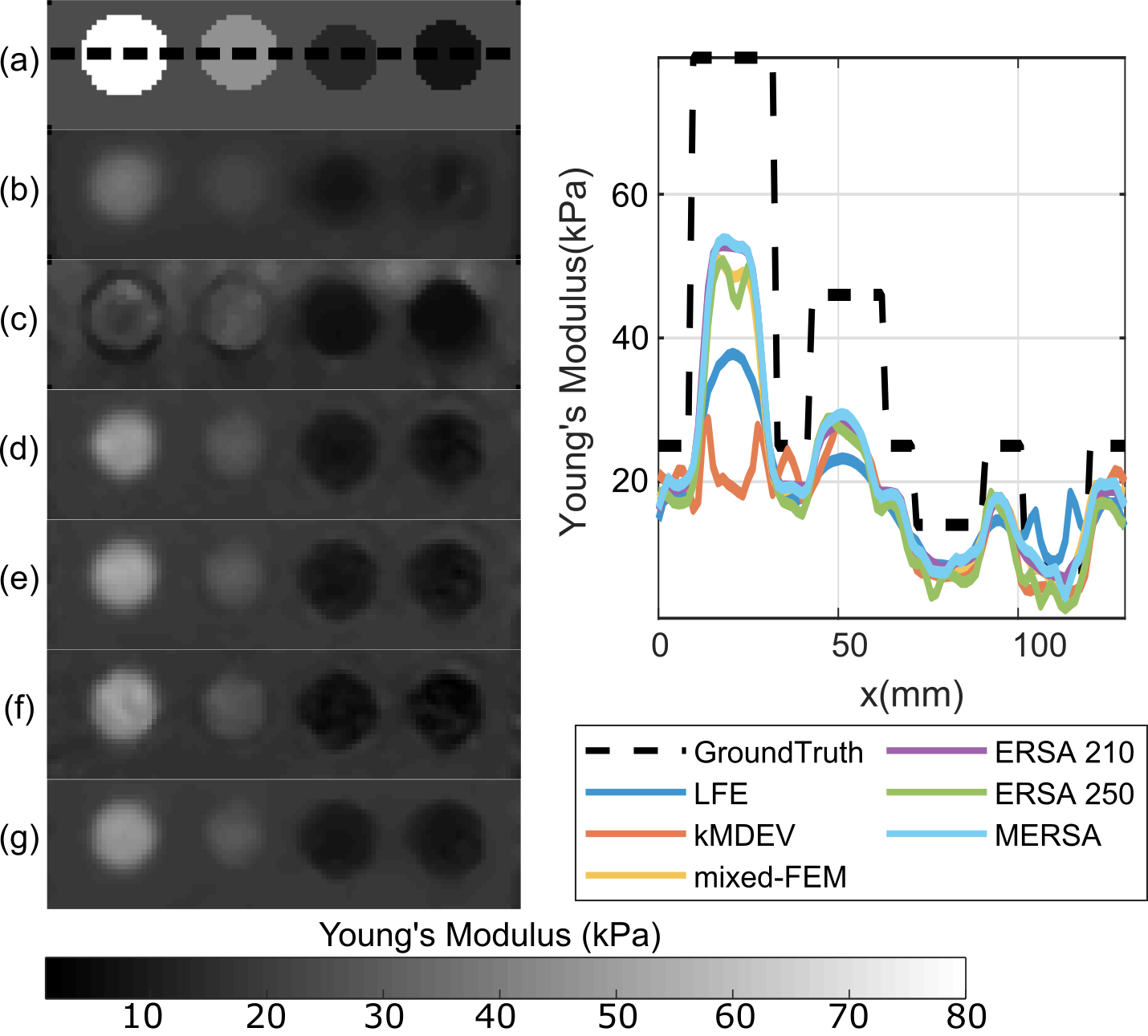}}
\caption{\TMISecondRevision{Left: Visual comparison of the CIRS~049 phantom mid-slice reconstructed with (b) LFE, (c) kMDEV, (d) Mixed-FEM, (e) ERSA at 210~Hz, (f) ERSA at 250~Hz, and (g) MERSA. 
(a) is the reference generated from segmenting the inclusions from the T2W image and assigning reported elasticities to the segmented regions. 
Right: line profile along the mid-line for all the reconstructed elasticity volumes with the corresponding references.} }
\label{CIRS049_MidSlice}
\end{figure}

\begin{comment}
\begin{figure*}[!t]
{\includegraphics[clip,width=1\linewidth]{./Figures/LiverHomoPhantom/Liver_Homo_CorrelationPlot.png}}
\caption{Correlation between reconstructed elasticity and reported elasticity from CIRS039 phantom for MERSA (a) and LFE (b). For comparison, we have also show the results from Philips Resoundant MRE in (c). All the methods underestimates the elasticity value, while Resoundant provides the closest value to the reported elasticity. However, MERSA shown the best correlation in terms of Pearson correlation coefficient and p-value.}
\label{CIRS_Homo_Plot}
\end{figure*}
\end{comment}
\TMIRevision{Fig. \ref{CIRS049_MidSlice} shows the mid-slice of the reconstructed elasticity for the CIRS~049 phantom obtained with LFE, kMDEV, Mixed-FEM and MERSA. 
% TS - repeat - We manually segmented each region and generated the reference ground truth image using the reported elasticity values from the high resolution T2W image.  
Qualitatively, MERSA resulted in well-defined inclusion boundaries with high contrast to the background, while there was some smoothing effect in LFE and Mixed-FEM. 
For quantitative evaluation, we have plotted the mid-line cutting through the inclusion in Fig. \ref{CIRS049_MidSlice}, which shows MERSA, ERSA, and Mixed-FEM to most closely follow the reference pattern.  
Table  \ref{table_CIRS049} confirms that MERSA provides the best CNR and ICC values. 
%{\bf $\leftarrow$ Footnote to Table IV on how the CIRS values were changed is unclear. Was the CIRS049 phantom tested with the Verasonics system? What does it mean "based on statistical comparison of CIRS compression test measurements...?}
}\par
\begin{comment}
\begin{table*}[!t]
%% increase table row spacing, adjust to taste
\begin{minipage}
\renewcommand{\arraystretch}{1.3}
% if using array.sty, it might be a good idea to tweak the value of
% \extrarowheight as needed to properly center the text within the cells
\caption{Comparison of reconstructed mean elasticity, standard deviation, CNR, and interclass correlation (ICC) with reported elasticity for different reconstruction algorithm for CIRS049 Phantom}
\label{table_CIRS049}
\centering
%% Some packages, such as MDW tools, offer better commands for making tables
%% than the plain LaTeX2e tabular which is used here.
\begin{tabular}{|c||c|c|c|c|c|}
\hline
Methods & LFE & DIFEM & HFEM & EPIQ &MERSA \\
\hline
Background (25) & $16.1$ & $16.8$& $16.3$ & $16.0$ & $15.4$\\
& 
\hline
Inc1 (8) & 7.5 & 10.3 & 11.3 & 5.6 & 3.2\\
\hline
Inc2 (14) & 8.3 & 7.3 & 5.8 & 8.2 & 6.3\\
\hline
Inc3 (46) & 22.8 & 24.2 & 25.7 & 26.8 & 21.4\\
\hline
Inc4 (79) & 34.8 & 38.3 & 39.6 & 46.2 & 39.2\\
\hline
\end{tabular}
\end{minipage}
\end{table*}
\end{comment}

\begin{table*}[!t]

\begin{threeparttable}
%% increase table row spacing, adjust to taste
\renewcommand{\arraystretch}{1.3}
% if using array.sty, it might be a good idea to tweak the value of
% \extrarowheight as needed to properly center the text within the cells
\caption{\TMISecondRevision{Comparison of reconstructed mean elasticity, standard deviation, CNR, and interclass correlation (ICC) with reported elasticity for different reconstruction algorithm for CIRS049 Phantom. The best result for CNR values and ICC are marked with bold. All the elasticity values are shown in~kPa.}}
\label{table_CIRS049}
\centering
%% Some packages, such as MDW tools, offer better commands for making tables
%% than the plain LaTeX2e tabular which is used here.
\begin{tabular}{|c||c|c|c|c|c|c|c|c|c|c|}
\hline
\multirow{2}{*}{\backslashbox{Methods}{Regions}}& Background & \multicolumn{2}{|c|}{Inclusion 1}&\multicolumn{2}{|c|}{Inclusion 2} &\multicolumn{2}{|c|}{Inclusion 3}& \multicolumn{2}{|c|}{Inclusion 4}& \multirow{2}{*}{ICC}\\
\cline{2-10}
& Mean & Mean& CNR&
Mean & CNR& Mean& CNR& Mean & CNR&\\
\hline
%Reported (Expected range)\tnote{a} & $25(22-28)$ & $8(7-9)$ &-& $14(12-16)$ &-& $46(40-52)$ &-& $79(70-88)$ &-&-\\
Reported expected range & $17.5-22$ & $5.5-7$ &-& $10-12.5$ &-& $32-41$&-& $56-71$&-&-\\
\hline
LFE \cite{Manduca1996} & $16.9 \pm 0.7$& $8.6 \pm 0.4$& $213$ &$9.1\pm 0.4$& $166$& $22.8\pm0.7$&$64.5$&$34.4\pm0.8$&$489$&$0.6484$\\
\hline
Mixed-FEM \cite{Honarvar2016} & $18.3\pm 0.8$& $6.2 \pm 1.0$& $162$&
$8.6\pm 0.4$& $320$& 
$27.5\pm0.9$&$107$&
$48.9\pm2.2$&$301$&
$0.8796$\\
\hline
kMDEV \cite{Tzschatzsch2016}&$18.3\pm 4.2$& $5.4 \pm 0.4$& $18.6$&$7.5\pm 0.4$& $9.86$& $23.5\pm2.1$&$2.5$&$26.6\pm2.8$&$5.5$&$0.5383$\\
\hline
%EPIQ &$16.1\pm 0.9$& $5.6 \pm 0.7$& $22.2$&$8.2\pm 1.1$& $17.8$& $26.8\pm1.4$&$19.3$&$46.1+0.9$&$\mathbf{30.5}$&$\mathbf{0.8619}$\\
%\hline
\TMISecondRevision{ERSA 210}&$18.6\pm0.5$&$8.2\pm1.4$&$95$&$9.1\pm1.0$&$129$&$28.1\pm1.0$&$127$&$50.7\pm1.7$&$612$&$0.8836$\\
\hline
\TMISecondRevision{ERSA 250}&$16.8\pm0.6$&$6.4\pm2.4$&$35$&$6.3\pm1.1$&$131$&$26.5\pm1.0$&$131$&$48.0\pm2.2$&$355$&$0.8796$\\
\hline
MERSA & $18.3\pm 0.4$& 
$7.5 \pm 0.7$& $\mathbf{316}$ &
$9.1\pm 0.3$& $\mathbf{394}$& $27.2\pm0.7$&$\mathbf{208}$&
$51.3\pm1.6$&$\mathbf{724}$&$\mathbf{0.8906}$\\
\hline
\end{tabular}
 %\begin{tablenotes}
  %  \item[b] Based on statistical comparison of CIRS compression test measurements and shear wave speed measurements using a Verasonics Vantage\texttrademark system.
  %\end{tablenotes}
\end{threeparttable}
\end{table*}

\begin{comment}
\begin{figure*}[!t]
{\includegraphics[clip,width=1\linewidth]{./Figures/CIRS_Phantom/New_Phantom_CIRS049.png}}
\caption{Reconstruction result for CIRS049 Phantom using different methods. The ground truth is generated by manual segmentation of the delineation of the inclusion in T2W image. }
\label{CIRS049_Phantom}
\end{figure*}
\end{comment}

\begin{comment}
\begin{figure}[!t]
\centering
{\includegraphics[clip,width=0.75\linewidth]{./Figures/CIRS_Phantom/Stats_Plot_new3.png}}
\caption{Distribution of mean elasticity for CIRS049 Phantom using different reconstructed elastography technique. The x-axis represents the reported elasticity  of the inclusion while y-axis represents the mean elasticity of the inclusion region. Standard deviation for each region with different reconstruction algorithm are shown with error bar. The The T2W mask shown in Fig. \ref{CIRS049_MidSlice} was used for extracting the regions for calculating the mean and standard deviation.}
\label{CIRS049_Stats}
\end{figure}
\end{comment}
\vspace{-0.1in}
\subsection{\emph{In vivo} Liver Elastography}
\label{subsec:: In-vivo Liver Elastography}
\begin{figure*}[!t]
\centering
{\includegraphics[clip,width=1\linewidth]{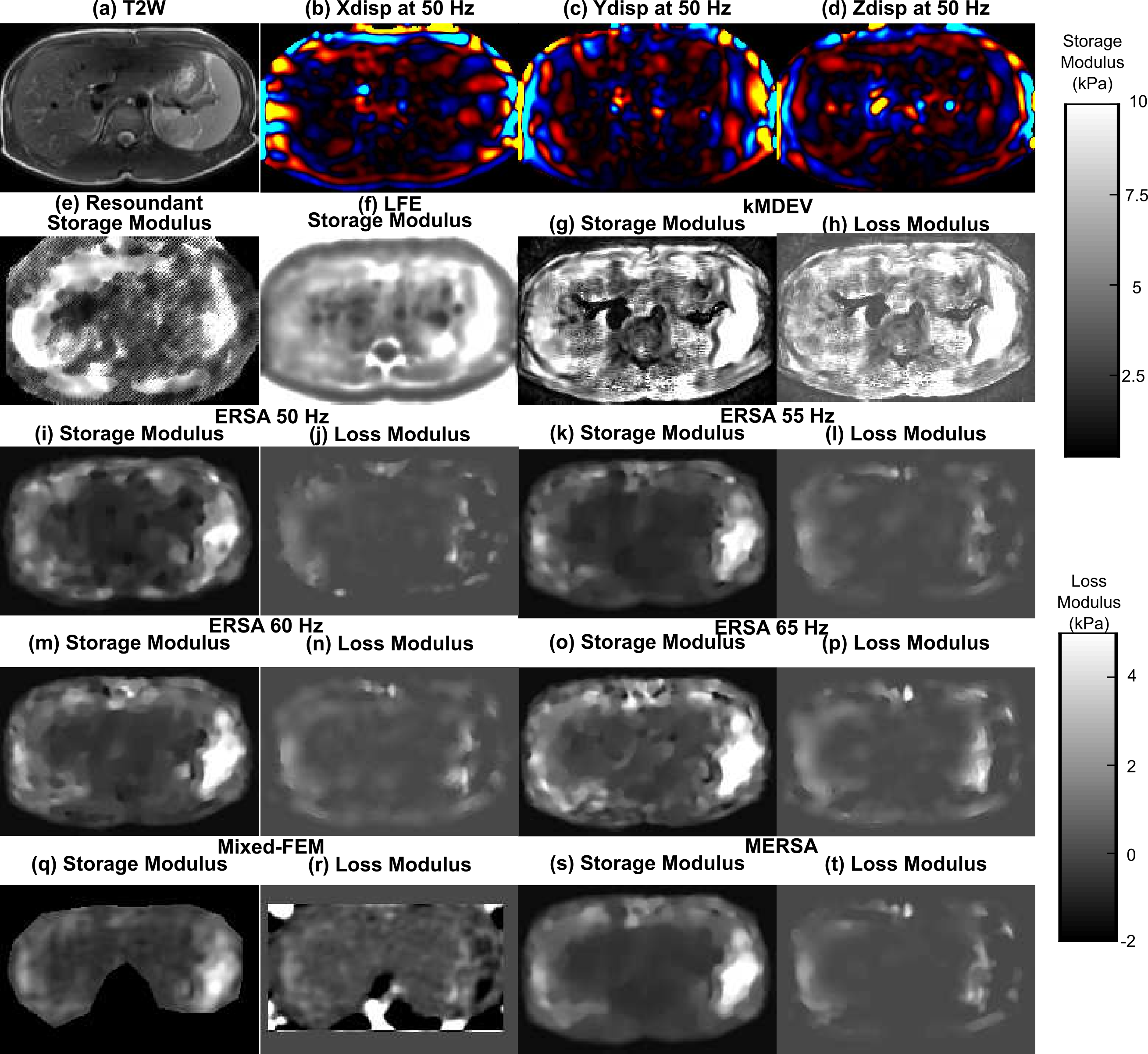}}
\caption{\TMISecondRevision{Visual comparison of Mid-plane of storage modulus and loss modulus for a volunteer using Resoundant MRE, LFE, kMDEV, Mixed-FEM, ERSA, and MERSA. LFE and Resoundant MRE do not provide a loss modulus map, so are not shown here. For reference, the T2W image and displacement phasor for 50 Hz frequency is shown in the top row.} }
\label{Liver_Images}
\end{figure*}

\TMIRevision{Fig. \ref{Liver_Images} shows the elasticity map reconstructed by different algorithms for a healthy volunteer. The leftmost image in the top row is the corresponding T2W image to give an anatomical reference. Also, we show the displacement phasor at 60Hz for all three directions in the top row. 
We compared the results with LFE~\cite{Manduca1996}, kMDEV~\cite{Tzschatzsch2016}, and Mixed-FEM~\cite{Honarvar2012}. 
It can be seen that the Mixed-FEM and MERSA provide similar results. 
We can also see from the results from ERSA, MERSA and Mixed-FEM, that the storage modulus and loss modulus estimates are very low at spatial locations where the wave amplitude is small inside the liver.} \par

\begin{figure}[!t]
\centering
{\includegraphics[clip,width=1\linewidth]{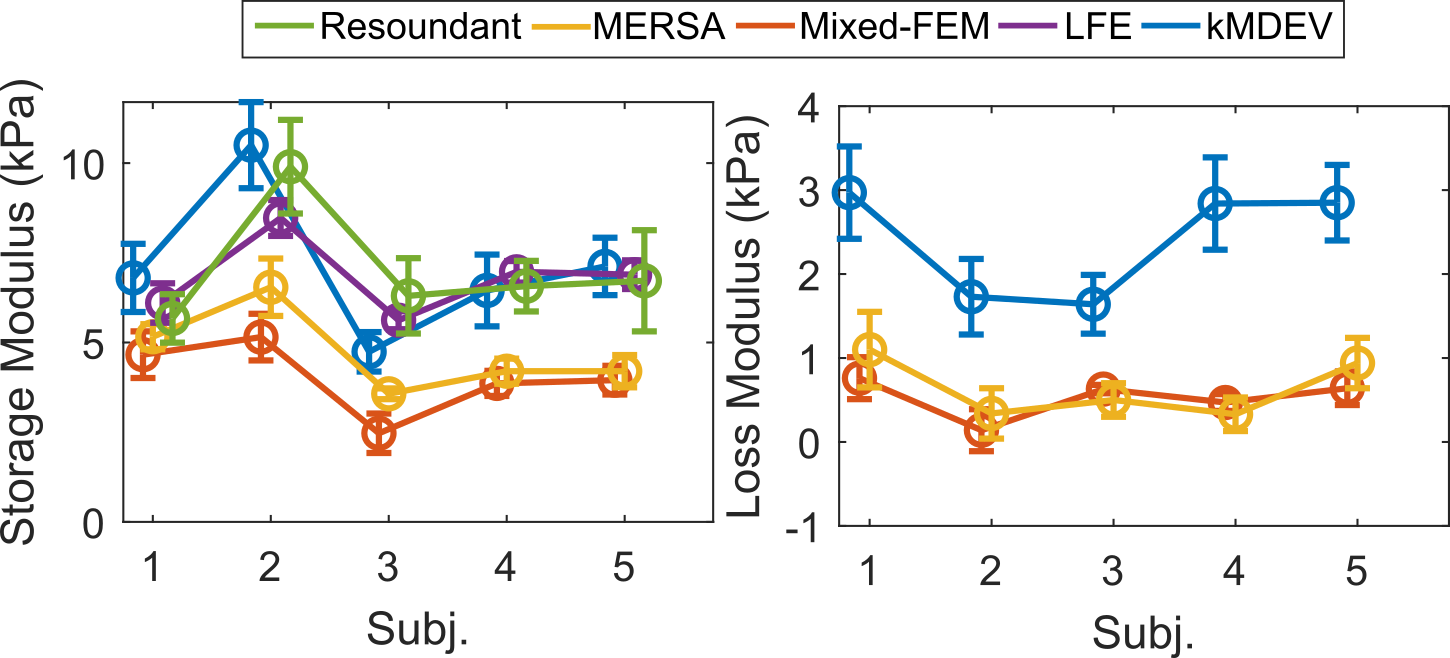}}
\caption{\TMISecondRevision{Comparison of mean storage modulus and loss modulus from different elasticity reconstruction methods. The error bar in each point represents the standard deviation. For calculating the mean and standard deviation of MERSA, Mixed-FEM, LFE, and kMDEV, region of interest (ROI) is manually drawn on the liver to discard vessels while ensuring sufficiently high displacement amplitude. For the Resoundant, ROI is drawn over the whole liver using MREView software on the scanner. The MREView software automatically rejects bad data points using the associated quality map.}
%{\bf $\leftarrow$ Storage repeated above. Please put these side by side , they don't have to be that large, and the legend can overlap the figure. TMI will not accept 14 pages. We will be very lucky if they accept 13. So it is important to reduce the length. I have cut quite a bit by reducing duplication and re-formatting but you need to fix the figures (Fig. 9, 5 and 3) yourself.}
}
\label{Liver_Stats}
\end{figure}

\TMIRevision{We compared the mean elasticity of the liver in five volunteers in Fig. \ref{Liver_Stats}. The region of interest is manually selected from the mid-slice, where  displacement amplitude is sufficiently high. All the methods follow a similar trend with subjects. However, there seems to be an underestimation of the mean storage modulus of Mixed-FEM and MERSA compared to LFE and kMDEV.}\par 

\section{Discussion and Conclusion}
\label{sec:: Discussion}
\TMIRevision{ Compared to finite difference based methods~\cite{Papazoglou2012}, FEM based methods only require first order differentiation of noisy measurements~\cite{Honarvar2012}. Furthermore, most of the finite difference-based methods require the local homogeneity assumption. 
Methods such as the heterogeneous multifrequency direct inversion (HMDI) do not require the local homogeneity assumption and have provided better CNR~\cite{barnhill2018heterogeneous}. 
FEM based methods also do not require local homogeneity assumption \cite{Honarvar2012}. 
Furthermore, FEM based methods such as the Mixed-FEM method take the hydrostatic pressure as an unknown~\cite{park2006shear, Honarvar2012}, and thereby do not require removing the divergence of displacement using noise amplifying methods such as curl filtering or high pass filtering. 
However, as Mixed-FEM methods have more unknowns than finite difference based methods, they require regularization. 
%{\bf Unclear to me why FEM methods require more unknowns than FD methods. Is this an accepted fact?}
For example, Mixed-FEM with sparsity regularization can reduce the number of unknowns and improve the reconstruction quality compared to curl-based finite difference methods and curl-based finite element methods~\cite{Honarvar2017comparison}.  However, as noted in previous studies, the results of Mixed-FEM direct inversion methods depend on the displacement quality and mesh density~\cite{Otesteanu2018}. 
This is also evident in the Mixed-FEM results for the homogeneous phantom given in Fig.~\ref{Homo_Plot2}. A significant increase in RMSE with noisy measurement is observed as $r_m$ value decreases. Fig~\ref{Hetero1Inc_Image} and Fig.~\ref{Hetero3Inc_Image} show a similar trend for detecting stiff inclusions, where noise causes underestimation of the inclusion elasticity. } \par
\vspace{-0.1in}
\label{Discussion:Para1}
\TMIRevision{Iterative methods can potentially provide more robustness to measurement quality, as the measurement is fitted with a model, which offers displacement noise filtering.
However, an iterative method based on the Mixed-FEM model suffers from higher instability than direct inversion methods~\cite{Honarvar2016, Otesteanu2018}.}\TMISecondRevision{ 
In our preliminary work~\cite{Mohammed2019}, bi-convex ADMM has shown improved stability,  convergence, and robustness for a iterative method with 2D shear wave equation model~\cite{Mohammed2019}.} { In ERSA and MERSA, we utilize a similar framework for a 3D wave constraint model with Mixed-FEM formulation, and note similar performance for numerical experiments for both single frequency and multi-frequency implementation. In contrast to the Mixed-FEM iterative model shown in~\cite{Honarvar2016}, our method improves the reconstruction quality significantly both in terms of accuracy~(Fig.~\ref{Homo_Plot2}) and contrast~(Fig.~\ref{Hetero1Inc_Plot}).
%{\bf $leftarrow$ I am not sure what you mean by the two previous sentences}
Furthermore, both ERSA and MERSA use wave-constraint, total variation denoising, and k-space sparsity filtering to filter out the noise from the measured displacements and elastogram reconstruction.  
The k-space sparsity prior works as an adaptive bandpass filter, reducing the measurement noise, which is generally dense and small in the k-space of the displacement~\cite{Barnhill2017}. 
%{\bf $leftarrow$ What does it mean "in small components"}
The wave equation constraint removes the noise component using displacement fitting. 
The box constraint on the shear modulus forces the fitted displacement to be constrained to a frequency band corresponding to the box constraint. 
Therefore it reduces the effect of high frequency sensor noise and low frequency bulk motions. 
As a result, both ERSA and MERSA can achieve much higher CNR than Mixed-FEM and LFE for stiff inclusions, as shown in Fig. \ref{Hetero1Inc_Image} and Fig. \ref{Hetero3Inc_Image}. } \par 
\label{Discussion:Para2}
\TMIRevision{Moreover, the use of multifrequency displacement measurements allowed MERSA to remove noise through averaging. 
ERSA is more susceptible to noise and discretization error, while MERSA can leverage complementary information to improve the reconstruction quality~\cite{Papazoglou2012, Tzschatzsch2016, Otesteanu2018}. 
However, the presence of local minima can increase in the solution space for multifrequency iterative methods, resulting in a stronger dependence on initialization~\cite{Otesteanu2018}. 
In MERSA, we have not observed a similar dependency with initialization. 
This is most likely due to the bi-convex nature of our formulation~\cite{Mohammed2019}, but is also helped by 
using the k-space prior that further constrains the solution space.} \par 
\TMIRevision{Results from the multi-inclusion phantom show that LFE provided a consistent result with noise; however, underestimation of the stiffer regions was observed. 
The effect of the noise was most prominent in Mixed-FEM.
Again, MERSA provided a smoother result by constraining the displacement solution space to be sparse in k-space, filtering the measurement noise. 
Thereby, MERSA provided the best delineation of the inclusion as evident from the line-profile in Fig. \ref{Hetero3Inc_Image}.} \par
\label{Discussion:Para3}

\TMIRevision{Both Mixed-FEM and MERSA find the complex stiffness modulus, where the imaginary part represents the loss modulus, and the real part represents the storage modulus. 
For the multi-inclusion numerical phantom with loss modulus, we see from Fig.~\ref{Hetero1Inc_Image} and Table~\ref{table_Hetero3Inc} that the Mixed-FEM method results in higher RMSE for the loss modulus than storage modulus even in the noiseless case. The difference in scaling between the real and imaginary part of the shear modulus might cause higher RMSE in the loss modulus. }
\TMISecondRevision{Fig. \ref{Hetero3Inc_Image} shows a significant artifact in the stiffest phantom for the 300~Hz implementation of ERSA and mixed-FEM. However, using multifrequency displacement data, MERSA provided the lowest $\text{RMSE}_\text{V}$.}  \par
\label{Discussion:Para4}
In our simulations and experiments, we focused on elastography reconstruction as a biomarker from both quantitative and qualitative perspectives. 
We experimented with elasticity values and inclusion sizes commonly occurring in human tissue with frequencies and vibration amplitude comparable to clinical settings. 
\TMIRevision{The numerical and experimental results show that MERSA can separate both soft and hard inclusions with lowest RMSE and highest CNR. Tumors are often harder than the surrounding tissue \cite{Sebastian2017_Book}, while degenerative diseases such as Alzheimer \cite{murphy2012magnetic} and dementia \cite{huston2016magnetic} can cause tissue to become softer.} 
Moreover, anatomical details were retained with better accuracy in MERSA. 
Anatomical details such as edges and boundaries are represented with high frequency sparse signals in the displacement pattern, often filtered out by the bandpass filtering. 
MERSA utilizes the elasticity estimate and wave constraint model to filter the Gaussian noise and compressional wave from the displacement pattern. 
The k-space sparsity prior diminishes dense noise without compromising anatomical details, increasing the image quality and thus the clinical value of quantitative elastography. \par
\label{Discussion:Para5}
\RobRevision{The result of the CIRS 049 phantom showed that qualitatively MERSA could delineate all the inclusions. 
However, the reconstructed values are underestimated with respect to the values reported by the manufacturer. 
MERSA showed similar reconstruction values as the other methods. 
We hypothesise that some underestimations may be due to the quasi-static nature of the compression test used for measuring the elasticity by CIRS and the dynamic nature in MRE, as CIRS uses a quasi-static compression test. 
In contrast, we have measured the elasticity from steady state harmonic waves at 210 and 250 Hz. 
\TMIRevision{However, considering a complex shear modulus that would account for an attenuation/viscosity component has led to negligible values for the imaginary part of the shear modulus \cite{fovargue2018robust}; further modelling is the subject of current research.} According to CIRS, a similar bias is also present between the CIRS reported values and transient elastography measurements for shear wave elasticity imaging (Table III-Adjusted expected range). It is also possible that the effect of the heterogeneous boundary and the refraction of the shear waves may cause a further discrepancy between the reported value and the measured value for the inclusions.} \par
\label{Discussion:Para6}
\TMIRevision{In the {\em in vivo} liver elastography, we see that LFE and kMDEV overestimate the elasticity compared to MERSA. 
One of the reasons can be bandpass filtering in LFE and the Gaussian smoothing filter in kMDEV. 
Particularly for LFE, we see the overestimation of the inclusion and background elasticity in Fig. \ref{Hetero3Inc_Image}. 
We can see a similar bias between kMDEV and MERSA for the CIRS039 phantom in Table \ref{table_CIRS039Homo}, where values from MERSA are closer to the values obtained by the QIBA protocol. 
Four out of the five volunteers we imaged had a mean elasticity from MERSA in the range of normal ($\leq 6~kPa$)\cite{mueller2020introduction}.}\par
\label{Discussion:Para7}
In terms of computational complexity, MERSA has lower computational complexity per iteration $\left(\left(N_wP+1\right)\left(\mathcal{O}\left(N^\frac{7}{3}\right)+\mathcal{O}\left(N^\frac{5}{3}\right)\right)\right)$ than the  conventional sub-zone based gradient descent method $\left(2\left(N_wP\right)\left(\mathcal{O}\left(N^\frac{7}{3}\right)+\mathcal{O}\left(N^\frac{5}{3}\right)\right)\right)$ \cite{Tan2017}. 
Here, we have assumed only the complexity for local displacement update and local elasticity update, as these steps dominated the computation. 
Moreover, similar to ERBA \cite{Mohammed2019}, MERSA can converge much faster to a good solution than other iterative methods and thus requires fewer iterations because of the use of bi-convex ADMM. 
Also, the sub-zone based iterative method requires higher overlapping between the sub-zones to reduce the blocking artifact. 
Using the global prior on elasticity in MERSA reduced the blocking artifacts and thus reduced the need for overlap. 
As a result, it is a faster iterative method, which could be made much faster using parallel computation by exploiting the distributed nature of ADMM algorithms \cite{Boyd2010}. \par
\label{Discussion:Para8}
There are several limitations of the presented tests. 
First, the number of volunteers was limited and was constrained to healthy livers. 
In future experiments, we will incorporate patients with different Fibrosis stages and liver diseases to validate the performance of the proposed method. 
Second, the effect of the regularization parameter and the region of interest was not assessed in this paper. 
In a clinical setting, the difficulty in selecting a regularization parameter may undermine image quality improvement. 
Therefore, incorporating an automated tuning mechanism for the regularization parameters and devising quality maps based on the reconstruction error may improve this work's clinical value. 
Third, we have not considered the effect of the poroelasticity of the vessels on elasticity measurement. 
We will address this issue  by introducing a more accurate FEM model in our framework in future studies. 
\TMIRevision{Fourth, the penetration of the vibration waves in the liver is limited. 
As a result, Mixed-FEM, ERSA, and MERSA produced valid elastograms only for a small region in the liver. 
This limitation is common to most methods. Addressing it may involve improvements to reconstruction methods, but also the use of multiple tissue exciters \cite{parker2017reverberant}. }%{\bf Please cite the work by Kevin Parker here}.
\TMISecondRevision{ We see that reconstructed elastogram from LFE and kMDEV provide more anatomical details than Mixed-FEM, ERSA, and MERSA. The use of directional filtering in LFE and kMDEV can provide attenuation compensation in the middle part of the liver, resulting in a wider coverage. However, the phase discontinuity in the organ boundaries can cause regions with artificial low stiffness. Furthermore, as the directional filtering can cause significant noise amplification, the results from LFE and kMDEV strongly depend on the smoothing filter. Nevertheless, the mixed-FEM based method are limited in providing attenuation compensation in the low displacement amplitude region and, therefore provide limited coverage of the liver.}  \TMISecondRevision{ Fifth, the voxel size is not isotropic due to constraint brought up by breath-hold time. However, simulations with anisotropic voxelsize ($1.5~mm\times1.5~mm\times4.5~mm$) with the viscoelasticit phantom have shown that the mean value of elasticity results change by less thant 5\% relative to the isotropic case ($1.5~mm\times1.5~mm\times1.5~mm$).} \TMIRevision{Lastly, no registration was applied between the scans from the different breath-holds in the liver experiment, which may cause errors in the elastograms. 
 Using the associated magnitude image to register each scan can be a feasible way to apply registration. 
 However, our initial effort in deforming the displacement pattern introduced more artifacts due to distortion of the wave pattern. 
 In the future, we will address the problem of registration by incorporating the registration as a prior for the reconstructed elastogram. }       \par
\label{Discussion:Para9}
In this work, we have presented the MERSA elasticity inversion algorithm based on the solution of a bi-convex wave constraint using ADMM and dual sparsity in displacement and stiffness. To reduce the computational complexity, we have introduced a sub-zone based method with a global sparsity prior on the elasticity. MERSA has shown excellent reconstruction performance in both numerical and phantom studies, with high accuracy and contrast. The experimental results showed it can provide highly detailed elasticity maps while keeping  excellent correlation with the other algorithms. \TMIRevision{Although this work was based on an incompressible mixed FEM model, it can be extended to other FEM models such as poroelastic or compressible viscoelastic models.} MERSA extends the iterative elasticity reconstruction and holds promise to provide high-quality fine resolution tissue elasticity and viscosity maps.
\label{Discussion:Para10}
\appendix
\section*{Multifrequency Extension}
\TMISecondRevision{In the Eqn. \ref{Optimization Model_MF}, $\mathbf{U}$, $\mathbf{V}$, $P$, $\mathbb{K}$, $\mathbb{M}$, $\mathbb{K}_p$, $\mathcal{R}_{U}$ and $\mathcal{R}_{p}$ are defined as:
\begin{eqnarray}
\mathbf{U} & = &\left[\begin{array}{ccc}
%\mathbf{u}_1^T & \mathbf{u}_2^T &  \hdots & \mathbf{u}_J^T \end{array}\right]^T
\mathbf{u}_1^T   \hdots & \mathbf{u}_J^T \end{array}\right]^T
\,\, ;  \,\,
\mathbf{V} =\left[\begin{array}{ccc}
\mathbf{v}_1^T  & \hdots &\mathbf{v}_J^T \end{array}\right]^T  \nonumber \\
\mathcal{R}_{U}\left(\mathbf{U}\right) &=&\norm{\hat{\mathbf{U}}}_1=\norm{\left[\begin{array}{cccc}
\hat{\mathbf{u}}_1^T &  \hdots & \hat{\mathbf{u}}_J^T \end{array}\right]^T}_1 \,\, , \nonumber \\
{P} &=& \left[\begin{array}{ccc} p_1^T  & \hdots & p_J^T \end{array}\right]^T
\,\, ; \,\,
\mathcal{R}_{P}\left(P\right) = \frac{\gamma_P}{2} \norm{\nabla P}_2^2 \nonumber \\
\mathbb{K}\left( \mu\right)&=&  \mathbb{I}_{J \times J} \otimes \mathbf{K} \left( \mu\right)
\,\,\, ; \,\,\,\,
\mathbb{K}_p =  \mathbb{I}_{J \times J} \otimes \mathbf{K}_p \nonumber \\
\mathbb{M}&=&  diag\left(\left[\begin{array}{ccc}
\omega_1^2 & \hdots & \omega_J^2 \end{array}\right]\right) \otimes \mathbf{M}
% \mathbb{K}_p&=  \mathbb{I}_{J \times J} \otimes \mathbf{K}_p \\
\end{eqnarray}
where $\mathbb{I}_{J \times J} $ is the  $J \times J$ identity matrix and $\otimes$ is the Kronecker product. We can also express the multifrequency 
equality constraint in linearized form for $\mu$ by reorganizing in the following way:
\begin{equation}
\mathbb{K}_{u}\, \mathbf{\mu}+\mathbb{K}_p P= 
\left[\begin{array}{c}
\mathbf{K}_{u}\left(\mathbf{u}_1\right) \\ \vdots\\ \mathbf{K}_{u}\left(\mathbf{u}_J\right) \end{array}\right]\mathbf{\mu}+\mathbb{K}_p P=\left[\begin{array}{c}
\mathbf{f}\left(\mathbf{u}_1\right) \\ \vdots\\ \mathbf{f}\left(\mathbf{u}_J\right) \\ \end{array}\right]\\ =\mathbb{F} \label{Linearized Equation_IPMF_Reorganized} \,\,\, 
\end{equation}
\vspace{-0.1in}
}
\section*{Optimization using Bi-convex ADMM}
\TMISecondRevision{The ADMM sub-problems are described next. We represent the estimate at ADMM iteration $k$ by a superscript.} 
\vspace{-0.1in}
\subsection*{$\nu_i$ Update: Local Elasticity Inversion with Hydrostatic Pressure}
\label{subsubsec:: LocalElasticityInversion}
\TMIRevision{With all other parameters fixed, (\ref{Optimization Model_AL}) is separable for $\nu_i$ and $Q_i$. Here, both $\nu_i$ and $Q_i$ are complex variables. Therefore, at ADMM iteration $k$,  we can find the local estimate of elasticity ${\nu}_i^{k+1}$ and $Q_i^{k+1}$ for the $i^{th}$ sub-zone by solving:}
\begin{align}
\label{Local Elasticity Update}
 \underset{\begin{subarray}{c}
	\nu_i \in \mathbb{C}^{M}\\
	Q_i \in \mathbb{C}^{MJ}\\
	\end{subarray}}{\min} &
 \,\,\,
\Bigg\{ \frac{\alpha_c}{2} \norm{\mathbb{K}_{u} \left( \mathbf{W}_i^{k}\right) \, \nu_i+ \mathbb{K}_{p}\,Q_i- \mathbb{F}\left(\mathbf{W}_i^{k}\right)+{\lambda_c}_i^k}^2
 +\nonumber\\
 &\frac{\alpha_\mu}{2} \norm{\nu_i^{k}-\mathbb{T}_i{\mu}^{k}+ \lambda_{\mu,i}^k}^2+ \frac{\gamma_p}{2}\norm{\nabla Q_i}^2 \Bigg\} 
\end{align}
Here, we utilized the bi-convex nature of the wave constraint through (\ref{Linearized Equation_IPMF_Reorganized}). This sub-problem is essentially an FEM based direct inversion of the local elasticity update from the current estimate of $\mathbf{W}_i^k$. The global estimate $\mu$ works as a reference elasticity for regularization, similarly to \cite{Honarvar2012, park2006shear}.  \vspace{-0.1in}
\subsection*{$\mu$ Update: Global Elasticity Update Step}
\label{subsubsec:: Update: Global Elasticity Update Step}
The second sub-problem in the ADMM loop is to find the global elasticity update $\mu$ from the current estimate of local elasticity updates while minimizing the TV prior. For this, we first calculated the global average $\overline{\nu}$ and $\overline{\lambda}_{\mu}$ over all the local estimates $\nu_i$ and the Lagrange multipliers $\lambda_{\mu,i}$, respectively \cite{Boyd2010}. Then, to find ${\mu}^{k+1}$ we solve:
\begin{align}
\underset{\begin{subarray}{c}
	\mu \in \mathbb{C}^{N}\\
	\end{subarray}}{\min} &
 \,\,\, \frac{\alpha_\mu}{2} \norm{\mu- \overline{\nu}^{k+1}-\overline{\lambda}_{\mu}^{k}}^2 +  \gamma_{\mu} \norm{\mu}_{TV} 
 \label{Elasticity_Denoise}
\end{align}
which is a TV denoising problem with noisy image $\overline{\nu}^{k+1}+\overline{\lambda}_{\mu}^{k}$, which we solve using a the algorithm from \cite{Beck2009}, where we have also incorporated a box constraint as a prior. \TMIRevision{As $\mu$ is a complex quantity, we apply TV denoising separately on the real and imaginary parts of the $\mu$}.  
\subsection*{$\mathbf{W}_i$ Update: Local Displacement Fitting}
\label{subsubsec: DisplacementFitting}
Similarly to the $\nu_i$ update, $\mathbf{W}_i$ is separable in (\ref{Optimization Model_AL}), allowing us to obtain 
the local displacement $\mathbf{W}_i$ by solving:
\begin{align}
\label{Local Displacement Update}
\underset{\begin{subarray}{l}
	\mathbf{W}_i \in \mathbb{C}^{3MP}\\
	\end{subarray}}{\min} &
 \,
\Bigg\{ \frac{\alpha_c}{2} \norm{\left[ \mathbb{K}_\mu \left( \nu_i^{k+1}\right)- \mathbb{M}\right]\,\mathbf{W}_i+ \mathbb{K}_P \, Q_i^{k+1}+ \lambda_{c,i}^{k}}^2 
 +\nonumber\\
 &\hspace{-0.6in}\frac{\rho}{2} \norm{ \mathbf{W}_i-\mathbb{S}_i\mathbf{V}}^2+\frac{\alpha_W}{2} \norm{FFT\left(\mathbf{W}_i\right)-\hat{\mathbf{W}}_i^{k}+ \lambda_{W,i}^k}^2 \Bigg\} 
\end{align}
This is similar to the least square solution of the forward problem with displacement boundary conditions, and is solved separately for each frequency exploiting its diagonal structure to significantly reduce its computational complexity. 
\vspace{-0.1in}
\subsection*{ $\hat{\mathbf{W}}_i$ Update: Displacement Regularization}
The fourth sub-problem involves soft-thresholding \cite{Boyd2010} of the $FFT\left(\mathbf{W}_i^{k+1}\right)+\lambda_{W,i}^{k}$ to find $\hat{\mathbf{W}}_i^{k+1}$:
\begin{align}
&\underset{\hat{\mathbf{W}_i}}{\min} \frac{\alpha_X}{2} \norm{FFT\left(\mathbf{W}_i^{k+1}\right)-\hat{\mathbf{W}}_i+ \lambda_{W,i}^{k}}^2
	 +\gamma_u\norm{\hat{\mathbf{W}}_i}_1  \label{Displacement_Denoise}
\end{align}
\vspace{-0.25in}
\subsection* {Dual Updates:}
In the last step, the scaled dual variable are updated based on the error on the current estimates \cite{Boyd2010}:
%\begin{subequations}
\begin{align}
	{\lambda_{c,i}}^{k+1}&= {\lambda_{c,i}}^{k}+ \left( \left[\mathbb{K}_\mu \left( \nu_{i}^{k+1}\right)- \mathbb{M}\right]\, \mathbf{W}_i^{k+1}+ \mathbb{K}_p\, Q_i^{k+1}\right) % \label{Dual_Variable_Update1} 
	\nonumber \\
    {\lambda_{W,i}}^{k+1}&= \lambda_{W,i}^{k}+ \left( FFT\left(\mathbf{W}_i^{k+1}\right)-\hat{\mathbf{W}}_i^{k+1}\right) %\label{Dual_Variable_Update3}	
    \nonumber \\
	{\lambda_{\mu,i}}^{k+1}&= {\lambda_{\mu,i}}^{k}+ \left( {z^{k+1}}- \mathbb{T}_i\,{\mu}^{k+1}\right)\label{Dual_Variable_Update4}
	\end{align}	
%\end{subequations}
The Lagrange multiplier accumulates the error in the wave constraint as well as the mismatch between the global elasticity estimate and local estimate. Therefore, with each iteration, the local estimate converges to a better fitting wave with higher conformity to the global elasticity distribution $\mu$.%
\if CLASSOPTIONcaptionsoff
  \newpage
\fi

\bibliographystyle{IEEEtran}
\bibliography{library_MERSA.bib}{}

% if you will not have a photo at all:
%\begin{IEEEbiographynophoto}{John Doe}
%Biography text here.
%\end{IEEEbiographynophoto}

% insert where needed to balance the two columns on the last page with
% biographies
%\newpage

%\begin{IEEEbiographynophoto}{Jane Doe}
%Biography text here.
%\end{IEEEbiographynophoto}
%
% You can push biographies down or up by placing
% a \vfill before or after them. The appropriate
% use of \vfill depends on what kind of text is
% on the last page and whether or not the columns
% are being equalized.

%\vfill

% Can be used to pull up biographies so that the bottom of the last one
% is flush with the other column.
%\enlargethispage{-5in}

% that's all folks
\end{document}